\documentclass[9pt,sigplan,screen,table]{acmart}

\usepackage[T1]{tipa}
\usepackage[tt=false, type1=true]{libertine}
\usepackage[libertine]{newtxmath}
\usepackage[shortcuts]{extdash}
\usepackage{seqsplit}

\usepackage{tikz,pgfplots}
\usetikzlibrary{arrows.meta}

\pgfplotsset{every axis/.append style={font=\sffamily\scriptsize}}
\newcommand*\circled[1]{\protect\tikz[baseline=(char.base)]{
            \protect\node[shape=circle,draw,inner sep=0.25pt] (char) {#1};}}

\usepackage{float}
\usepackage{tipa}
\usepackage{graphicx}
\usepackage{balance}
\usepackage{stfloats}
\usepackage[caption=false,font=footnotesize,labelfont=sf,textfont=sf]{subfig}
\usepackage{booktabs}
\usepackage[symbol]{footmisc}
\usepackage{makecell}
\usepackage{mdframed}
\usepackage{listings}
\usepackage{bm}
\usepackage{enumitem}
\usepackage{multirow}
\usepackage{nicefrac}
\usepackage{amsmath}

\usepackage{amssymb}
\usepackage{booktabs}
\usepackage{todonotes}
\usepackage{arydshln}
\usepackage{booktabs,xcolor,siunitx}

\usepackage{xcolor}
\usepackage{tcolorbox}
\definecolor{lightgrey}{HTML}{D3D3D3}
\definecolor{commentblue}{HTML}{155196}
\definecolor{emphgreen}{HTML}{4F6900}
\definecolor{emphpink}{HTML}{b5283c}
\definecolor{backgrey}{HTML}{E8E8E8}
\definecolor{backblue}{HTML}{EEF1F7}

\def\BibTeX{{\rm B\kern-.05em{\sc i\kern-.025em b}\kern-.08emT\kern-.1667em\lower.7ex\hbox{E}\kern-.125emX}}
    
\usepackage{color, colortbl}
\usepackage{soul}
\usepackage{xcolor}
\usepackage{tcolorbox}
\definecolor{lightgrey}{HTML}{E8E8E8}
\definecolor{lightgrey2}{HTML}{E8E8E8}
\definecolor{ForestGreen}{rgb}{0.0, 0.27, 0.13}
\newcommand{\ctext}[1]{{\sethlcolor{lightgrey2}\hl{#1}}}

\setcopyright{none}
\renewcommand\footnotetextcopyrightpermission[1]{}

\emergencystretch=\maxdimen
\begin{document}

\begin{titlepage}
  \mbox{}\\{\Large \textbf{IEEE Copyright Notice}}
  \newline\newline\newline\newline
  \textcopyright~2021 IEEE. Personal use of this material is permitted.
  Permission from IEEE must be obtained for all other uses, in any current
  or future media, including reprinting/republishing this material for
  advertising or promotional purposes, creating new collective works, for
  resale or redistribution to servers or lists, or reuse of any copyrighted
  component of this work in other works.
  \newline\newline\newline\newline
      {\large Extended version of a paper accepted for publication in: Proceedings of the 28th IEEE
        International Conference on High Performance Computing, Data, and Analytics (HiPC), December 17-18, 2021}
\end{titlepage}

\title[JACC: An OpenACC Runtime Framework with Kernel-Level and Multi-GPU Parallelization]{JACC: An OpenACC Runtime Framework\\ with Kernel-Level and Multi-GPU Parallelization}

\author{Kazuaki Matsumura, Simon Garcia De Gonzalo, Antonio J. Peña}
\affiliation{\hspace*{-7pt}\mbox{Barcelona Supercomputing Center (BSC)}}
\email{{kazuaki.matsumura, simon.garcia, antonio.pena}@bsc.es}

\renewcommand{\shortauthors}{K. Matsumura, S. G. De Gonzalo, A. J. Peña}

\setlength{\textfloatsep}{10pt}
\setlength{\dbltextfloatsep}{10pt}
\setlength{\abovedisplayskip}{0pt}
\setlength{\belowdisplayskip}{6pt}
\setlength{\abovedisplayshortskip}{0pt}
\setlength{\belowdisplayshortskip}{6pt}

\begin{abstract}
The rapid development in computing technology has paved the way for directive-based programming models towards a principal role in maintaining software portability of performance-critical applications. Efforts on such models involve a least engineering cost for enabling computational acceleration on multiple architec- tures while programmers are only required to add meta informa- tion upon sequential code. Optimizations for obtaining the best possible efficiency, however, are often challenging. The insertions of directives by the programmer can lead to side-effects that limit the available compiler optimization possible, which could result in performance degradation. This is exacerbated when targeting multi-GPU systems, as pragmas do not automatically adapt to such systems, and require expensive and time consuming code adjust- ment by programmers.

This paper introduces JACC, an OpenACC runtime framework which enables the dynamic extension of OpenACC programs by serving as a transparent layer between the program and the compiler.
We add a versatile code-translation method for multi\-/device utilization
by which manually-optimized applications can be distributed automatically
while keeping original code structure and parallelism.
We show in some cases nearly linear scaling on the part of kernel execution
with the NVIDIA V100 GPUs.
While adaptively using multi-GPUs,
the resulting performance improve- ments amortize the latency of GPU-to-GPU communications.
\end{abstract}


\maketitle
\pagestyle{plain}
\section{Introduction}\label{sec:introduction}

Designing and building supercomputers is a complex task in the field of high-performance computing (HPC). The hardware, mid- dleware and algorithms need to effectively collaborate to achieve ideal results for massive and practical problems. To facilitate the easy usage of supercomputers, compiler technologies have been developed with highly automated program optimizations that use domain-specific knowledge and understandings of target architec- tures~\cite{10.5555/572937}.

The adoption of newer technologies brings newer challenges.
Currently, supercomputers employ various kinds of accelerators in the range from SIMD units available on CPUs to discrete devices such as GPUs, ASICs and Intel Xeon Phis~\cite{top500}.
Those hetero- geneous systems tend to force a certain amount of engineering efforts on application developers to establish an adequate distri- bution of program execution and attain
performance for practical use.

Directive-based programming has been employed for enabling accelerator use,
while replacing vendor-specific coding with direc- tive insertion.
Keeping software portability
with minimum engi- neering efforts upon sequential code,
OpenACC and OpenMP are now widely used for accelerator programming~\cite{openacc, openmp}. However, pursuing ideal performance is often challenging.
The insertion of directives by the programmers results in compilation side-effects that lead to less program-characteristics exposure for compila- tion~\cite{10.1109/CCGrid.2013.12,10.1145/3110355.3110356,khalilov2021performance}; thus,
programmers aiming at better efficiency are forced to reshape their code merely for adjusting to the environ- ment such as compilers, software stacks and heterogeneous architecture.
Moreover, to follow the program modification,
addi- tional runtime parameters are often introduced
for each program segment.
Therefore, managing rewritten code is far from clear
regardless of the complexity of transformation.
Specifically, the parallelism among kernels is rarely addressed
and the multi-device utilization is basically dismissed
due to the little usability of data dependency information.

To explore new optimization opportunities,
this paper extends OpenACC to hide code redundancy
of optimization
behind the runtime system in order to facilitate compiler development.
While requiring no modification on original programs,
our framework JACC\footnote{Available at \url{http://github.com/epeec/JACC}.} provides an environment for dynamic analysis, reschedul- ing and distribution of execution along with on-the-fly kernel specialization by wrapping up existing OpenACC compilers.
Al- though other directive-based programming work develops ded- icated runtime systems for specific optimization~\cite{10.1145/2792745.2792783,10.1007/978-3-319-69953-0_7,10.1145/3178487.3178492},
our framework integrates the compilation and runtime phases
so as to utilize both aspects for additional efforts especially aiming at ex- ploiting parallelism.

Additionally,
we 
address multi-GPU work distribution%
, while
considering the high memory latency of GPUs.
To accomplish this, we add
a novel code-translation technique named {\it predicated-based filtering}
to automate multi-device use.
We never split loop ranges nor introduce fine dependency analysis, but
divide data ranges to be updated on each device.
This idea allows to distribute highly-tuned code
without changing code structure nor parallelism.
Our contributions are as follows:

\begin{itemize}[topsep=0.5em, leftmargin=0.15in,rightmargin=0.05in, itemsep=0.5em]
\item We create JACC, an OpenACC framework which facilitates various dynamic features including runtime data analysis and compilation. %
JACC enables kernel-level parallelization through an asynchronous mechanism.

\item We %
propose and describe a new multi-GPU kernel distribution method %
by leveraging JACC for complex applications. Our pro- posed technique is successful at multi-GPU execution on ap- plication that previous work fails to offload to multiple devices. 

\item %
We propose and describe an adaptive algorithm that automat- ically determines %
the adequate kernels for multi-GPU execution. 
Our adaptive algorithm considers 
the bandwidth of the peer-to-peer communication between GPUs 
when selecting kernels for multi-GPU execution.

\item We evaluate all our contributions by using two different OpenACC compilers on a NVIDIA V100 multi-GPU system. We show that 
using our proposed methods and techniques available through JACC, we are able to achieve nearly linear scaling %
when excluding the latency of communication. Addi- tionally, we are able to
successfully improve the performance of a variety of manually-tuned NAS Parallel Benchmarks.
\end{itemize}

The rest of the paper is structured as follows.
Section~\ref{sec:background} discusses the recent trend of GPUs alongside a brief overview 
of OpenACC.
Section~\ref{sec:jacc} introduces our runtime system with basic extensions.
In Section~\ref{sec:multigpu}, predicate-based filtering is described.
Section~\ref{sec:method} shows our experimental methodology.
Section~\ref{sec:results} evaluates our proposed technique on the state-of-the-art hardware.
Section~\ref{sec:related} discusses related work.
Finally, Section~\ref{sec:conclusion} concludes this paper.

\section{Background}\label{sec:background}
\subsection{GPUs}

\begin{figure}[b]
  \centering
  \hspace*{-0.15cm}
  \includegraphics[width=0.49\textwidth]{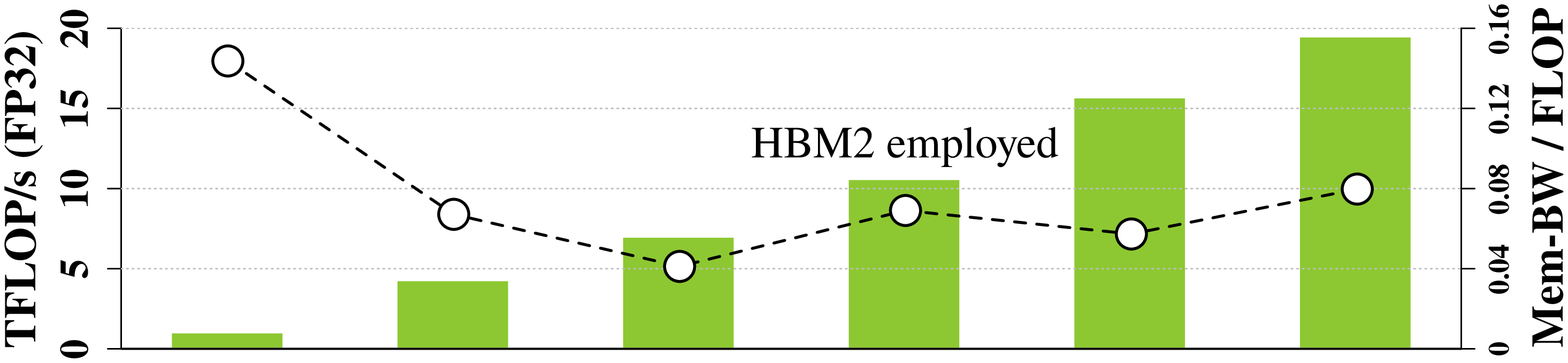}
  \hspace*{-0.15cm}
  \includegraphics[width=0.49\textwidth]{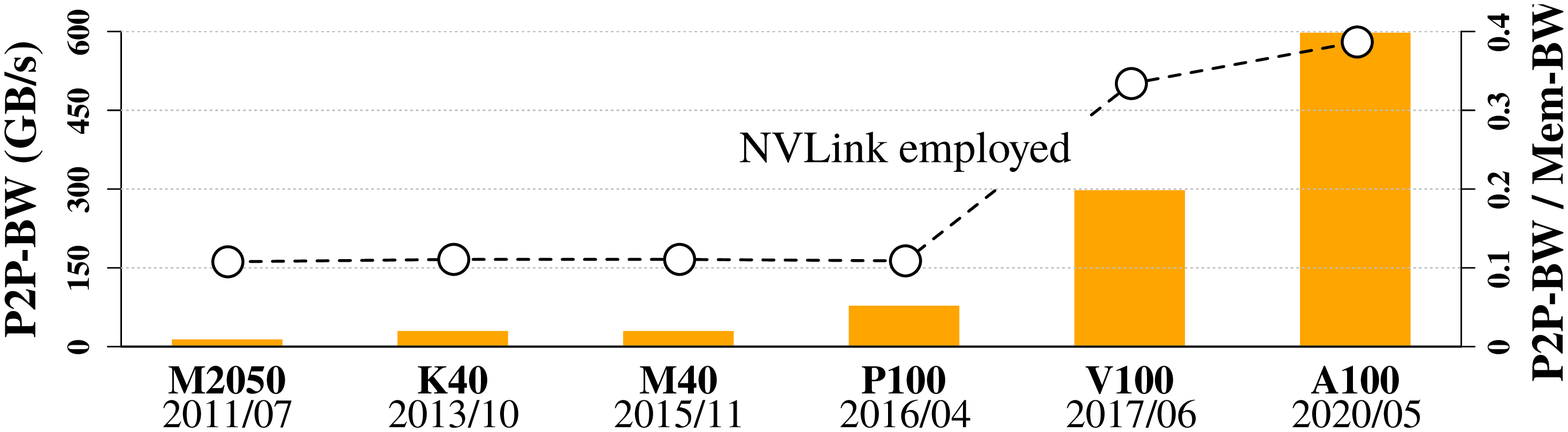}
  \vspace{-20pt}
  \caption{Performance changes of recent GPUs. The bars show peak performance in TFLOP/s (top) and peak GPU-to-GPU bidirectional bandwidth in GB/s. The ratio to the memory bandwidth is shown by the line plot.}
  \label{fig:trend}
\end{figure}

Graphics Processing Units (GPUs) originated as graphic processors supporting massive parallelism.
In contrast to CPUs,
which are used for general computations and system management,
GPUs are usually dedicated to those parts of applications featuring a high degree of parallelism.
Since GPUs can hide the latency of memory requests
by overlapping it with execution,
the peak bandwidth of GPUs is significantly larger than that of CPUs.
Furthermore, the hierarchical memory system including multi-level caches,
shared memory and registers, allows exploiting spatial and temporal locality.
We show the recent performance changes of NVIDIA GPUs~\cite{m2050, k40, m40, p100, v100, a100} in Fig.~\ref{fig:trend}.
Until P100, which began employing the second generation of High Bandwidth Memory (HBM2),
the performance gap between memory bandwidth and computational throughput had been growing larger generation after generation.
Therefore, memory accesses tend to be the performance bottleneck of applications.
Currently,
the NVLink interconnect~\cite{nvlink} is enhancing
peer-to-peer communication among multiple GPUs.
The NVLink bandwidth
allows
not only compute-intensive but memory-intensive applications
to utilize several devices.
Notably, Unified Memory (UM) realizes
data accesses that automatically solve data coherence
in the entire system while performing peer-to-peer communications.
However, frequent page faults cause
heavy performance degradation~\cite{10.1145/3295500.3356141};
hence, users are basically required to avoid data sharing
or perform explicit memory copies among GPUs.

GPU programs,
which are ordinarily created leveraging the CUDA programming model~\cite{cuda},
have to divide the execution into thread-blocks.
Each thread-block consists of the GPU's smallest unit of parallelism called threads.
In CUDA, the user programs each thread's behavior in one kernel.
The behavior is defined in an extended form of the C language
and controlled in accordance with thread id and thread-block id.
One thread has its own registers,
threads in one thread-block hold their own shared memory,
and users are responsible for setting their configurations
to obtain correct results.
The on-chip resource use decides processor occupancy, which limits final efficiency together with instruction-level parallelism (ILP) and synchronization.

\subsection{OpenACC}

Directive-based programming models have become common in HPC.
Application developers can take them for granted in most supercomputing environments equipped with modern accelerators such as GPUs or FPGAs.
Programmers benefit from their ability to target the same code easily to different possible accelerator architectures.
OpenACC~\cite{openacc} offers compiler directives to program accelerators
in existing languages.
Without introducing vendor\-/specific languages such as CUDA,
users are allowed to parallelize their code
and rely on the compiler for generating device-specific application code.
Listing~\ref{acc-sample} shows an OpenACC code written in C that
updates array x with the multiplication of array y.

\begin{figure}[h]
\vspace*{0pt}
\begin{lstlisting}[caption={Accelerator programming in OpenACC}, label={acc-sample}]
_*#pragma acc*_ data copyout(x[0:N]) present(y)
_*#pragma acc*_ parallel loop
for(int i=0; i<N; i++) x[i] = y[i] * y[i];
\end{lstlisting}
\vspace*{-10pt}
\end{figure}

The components of OpenACC are made of kernels and routines.
An OpenACC kernel is the unit of program execution on ac- celerators
to be launched with specified parallelism (consisting of \texttt{gang}/\texttt{worker}/\texttt{vector}).
Since the host code is executed on CPUs, kernel execution can be asynchronous to
CPU execution and multiple kernels can be simultaneously run
on the same device.
The environment for kernel execution,
such as device setting, data copies to/from devices and synchronous behavior,
can be con- trolled by OpenACC routines.

OpenACC directives are provided for specifying code segments
as kernels or defining data on devices along with several options (Lines 1-2 of Listing~\ref{acc-sample}).
Although data-related directives can be replaced by routine calls,
OpenACC kernels have to be embedded on original source files with directives
to be converted to device-specific code at compile time.
Therefore, additional code segments have to be put in place along with additional variables in order to be calculated from various dynamic parameters at runtime.
Listing~\ref{acc-multi} shows the example of multi-device utilization in OpenACC with asynchronous execution to each other devices.
Additional code segments surround the kernel,
calling an OpenACC routine to switch devices (Line 2 of Listing~\ref{acc-multi}) and setting variables to divide the loop execution (Lines 3-5 of the same figure).
In real applications, many other factors such as runtime information and device-to-device communication are concerned; hence, in-situ kernel declarations bring additional complexities to directive-based software development. %
Also, OpenACC kernels are statically declared;
thus, for a complex dynamic application the programmer is required to prepare adjusted kernels before compilation, regardless of whether run- time information is available.

\begin{figure}[t]
\vspace*{5pt}
  \centering
  \begin{lstlisting}[caption={Multi-device use in OpenACC}, label={acc-multi}]
for (int d = 0; d < NUM_DEVICES; d++) {
  acc_set_device_num(d, 0);
  int length = N/NUM_DEVICES;
  int init   = length * d;
  int until  = length * (d + 1);
  _*#pragma acc*_ data copyout(x[init:length])\
                   present(y) async(d)
  _*#pragma acc*_ parallel loop async(d)
  for(int i=init; i<until; i++)
    x[i] = y[i] * y[i];
}
\end{lstlisting}
\vspace*{-10pt}
\end{figure}

\section{JACC: Runtime-Extended OpenACC}\label{sec:jacc}

Our work introduces dynamic analysis and
compilation to OpenACC directive-based programming,
allowing further efforts on optimization at runtime.
All the components of OpenACC, here, are provided as
runtime routines leveraging existing compilers.
By transforming directives into a sequence of routine calls,
OpenACC compilers can enable on-the-fly features such as kernel special- ization and load-balancing.

\subsection{JACC}

We build JACC, a just-in-time compilation system for OpenACC,
in which input directives are replaced with runtime routines.
JACC hides every OpenACC feature behind a %
runtime library
to cushion dependency to specific compilers.
Once a kernel is compiled for the first time, 
its device code is cached to be reused for subsequent launches.
Even though JACC is developed upon existing compilers,
it allows calling of CUDA routines and kernels through its library.
Listing~\ref{jacc-sample} shows the converted code of Listing~\ref{acc-sample} to
call runtime routines. First, combined directives (e.g. \texttt{parallel loop} of Line 2 of Listing~\ref{acc-sample}) are decomposed into three basic directives of \texttt{parallel}, \texttt{loop} and \texttt{data}. Then, for each directive, JACC inserts corresponding routines that are implemented in its library, shown in Listing~\ref{jacc-sample} (Lines~2,~5~and~12).

\begin{figure}[h]
  \centering
  \begin{lstlisting}[caption={Converted code by JACC (arguments omitted)}, label={jacc-sample}]
/* Entry of #pragma acc data */
[*jacc_create*](x, N * sizeof(float));

/* #pragma acc parallel loop */
[*jacc_kernel_push*](
  "_*#pragma*_ acc parallel present(x, y)\n"
  "_*#pragma*_ acc loop\n"
  "for(int i=0; i<N ; i ++) /* ... */",
   /* args */, /* flags */);

/* Exit of #pragma acc data */
[*jacc_copyout*](x, N * sizeof(float));
\end{lstlisting}
\vspace*{-10pt}
\end{figure}

During program execution, JACC data-related routines that wrap OpenACC routines (Lines 2 and 12 of Listing~\ref{jacc-sample}) assume the roles of the original directives.
The routine {\bf\color{emphpink}\texttt{jacc\_kernel\_push}} launches kernel execution
while
accepting source code in a string with arguments that hold runtime information (Lines 5-9 of the same figure).
It should be noted that the \texttt{loop} directive is used for marking parallelism;
therefore, the directive is kept in kernel strings.
When the routine finds no compiled kernel for given source code
or needs to update existing kernels, function code (Listing~\ref{jacc-sample2}) is generated
to emit device code by a specified compiler
and to have additional arguments for code extension.
After linked dynamically, this function is called through a foreign function interface (FFI) for passing arbitrary arguments.
JACC's library for each routine is extended to collect runtime information
and support dynamic features.

\begin{figure}[h]
\vspace*{-5pt}
  \centering
  \begin{lstlisting}[caption={Generated code for kernel launch (formatted)}, label={jacc-sample2}]
void kernel0(float *x,
             float *y, size_t N) {
  _*#pragma acc*_ parallel present(x, y)
  _*#pragma acc*_ loop
  for(int i=0; i<N; i++) x[i] = y[i] * y[i];
}
\end{lstlisting}
\vspace*{-10pt}
\vspace*{-6pt}
\end{figure}

\begin{figure*}[h]
  \vspace*{0.2cm}
  \centering
  \hspace*{-0.15cm}
  \includegraphics[width=0.98\textwidth]{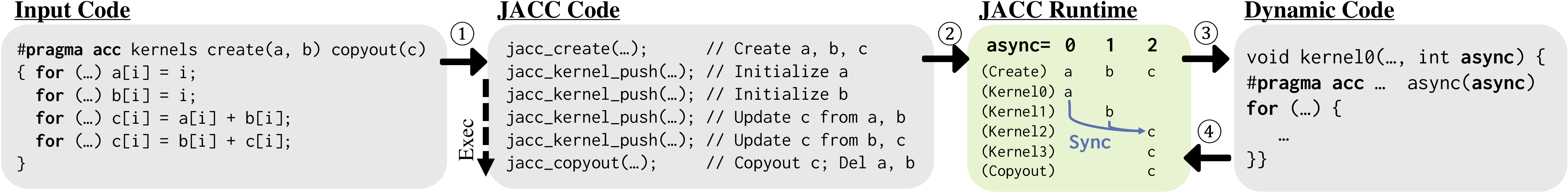}
  \vspace{-10pt}
  \caption{Execution flow of automated asynchronous execution. First, the JACC code is generated from the input \circled{1}. Following the execution, the runtime manages data declarations and checks data dependency among asynchronous queues based on data ranges of actual kernel arguments \circled{2}. Dynamic code is created from kernel code to be executed on a destined queue based on the kernel argument \texttt{\bf async} \circled{3} and synchronization is performed with the host if necessary \circled{4}.}
  \label{fig:async}
\end{figure*}

\subsection{Basic Extension}
\label{subsec:extension}
OpenACC is a high-level programming model designed for accelerator abstraction, which accepts program optimizations through both direct APIs and program modification in the base language (e.g. C, C++, Fortran). Whereas the OpenACC APIs are added explicitly to a program as directive clauses with a specific intent~\cite{10.5555/3433701.3433831,10.1145/2792745.2792783}, the intent and effect of base-language modifications to a program are implicit. There exist a large body of work that studies the implications of base-language modification on OpenACC compilation~\cite{hoshino, 10.1145/3243176.3243196, 7573861, 8672231}. JACC works as a runtime solution for dynamic optimization for both the OpenACC APIs and base-language program modification. Because JACC's ability to handle both types of optimizations,
it can automatically overlap kernel execution, and thus achieve inter-kernel parallelism. Also, additional
on-the-fly kernel specialization using runtime infor- mation extracted from profiling results for better resource use and intra-kernel parallelism are possible.

\subsubsection{Automated Asynchronous Execution}

Since JACC automatically provides a function interface for each OpenACC kernel, additional runtime information needed for extended execution shown in lines 2-5 of Listing~\ref{acc-multi}, can be passed through as arguments to the JACC function interface without the need to generate %
redundant code snippets. Instead of compiler\-/generated code, which is hard to manage as global program information, JACC's runtime calculates the required arguments on its own runtime environment and then proceeds to call the kernel functions using them. The benefit of this approach is that infor- mation across multiple execution instances can be easily shared for further optimizations. Moreover, JACC provides the ability to update kernels dynamically with additional runtime information after program compilation is invoked, thus, providing a straight- forward mechanism for runtime extensions of original kernel declaration.

For asynchronous execution, we automatically
overlap kernel launches and data operations with each other as well as host execution.
Each JACC routine has the ability %
to track array references, and if data dependencies are encountered across two or more routines, JACC will schedule them
in the same asynchronous queue.
When multiple queues are concerned,
synchronous operations are
performed only among those queues that require them,
while skipping redundant synchronization on already solved dependencies. JACC achieves this by maintaining
timestamps of data accesses and the most recent synchronization among queues.
If there is no data dependency to prior execution, the least recently used queue is selected.
We guarantee original code semantics
by obligatory synchronization that is performed immediately after
kernel execution that deals with array writes to undefined data regions and explicit variable updates such as reduction.
Data ranges linked to given pointers are tracked through JACC's runtime routines,
and managed in a red-black tree as OpenACC compilers do to accept any address of declared data~\cite{wolfe2018openacc}.

Fig.~\ref{fig:async} shows JACC's
automatic asynchronous code optimization and execution flow.
During execution by the JACC's runtime (Step 2 of Fig.~\ref{fig:async}), the synchronization between queues is performed. For example, before 
\texttt{Kernel2} execution is allowed, a synchronization call is performed to wait for the updated arrays \texttt{a} and \texttt{b}. However, for \texttt{Kernel3}, the dependency on \texttt{b} is already solved by the previous synchronization, thus the execution does not wait for other queues.
For the overlapping execution, during the dynamic code generation  (Step 3 of Fig.~\ref{fig:async}), the OpenACC clause \texttt{async} is set for each kernel declaration, and
asynchronous execution queues are selected.

\subsubsection{Kernel Specialization}
\label{subsubsec:kernel-spec}

\begin{figure}[t]
  \vspace*{0.2cm}
  \centering
  \hspace*{-0.15cm}
  \includegraphics[width=0.49\textwidth]{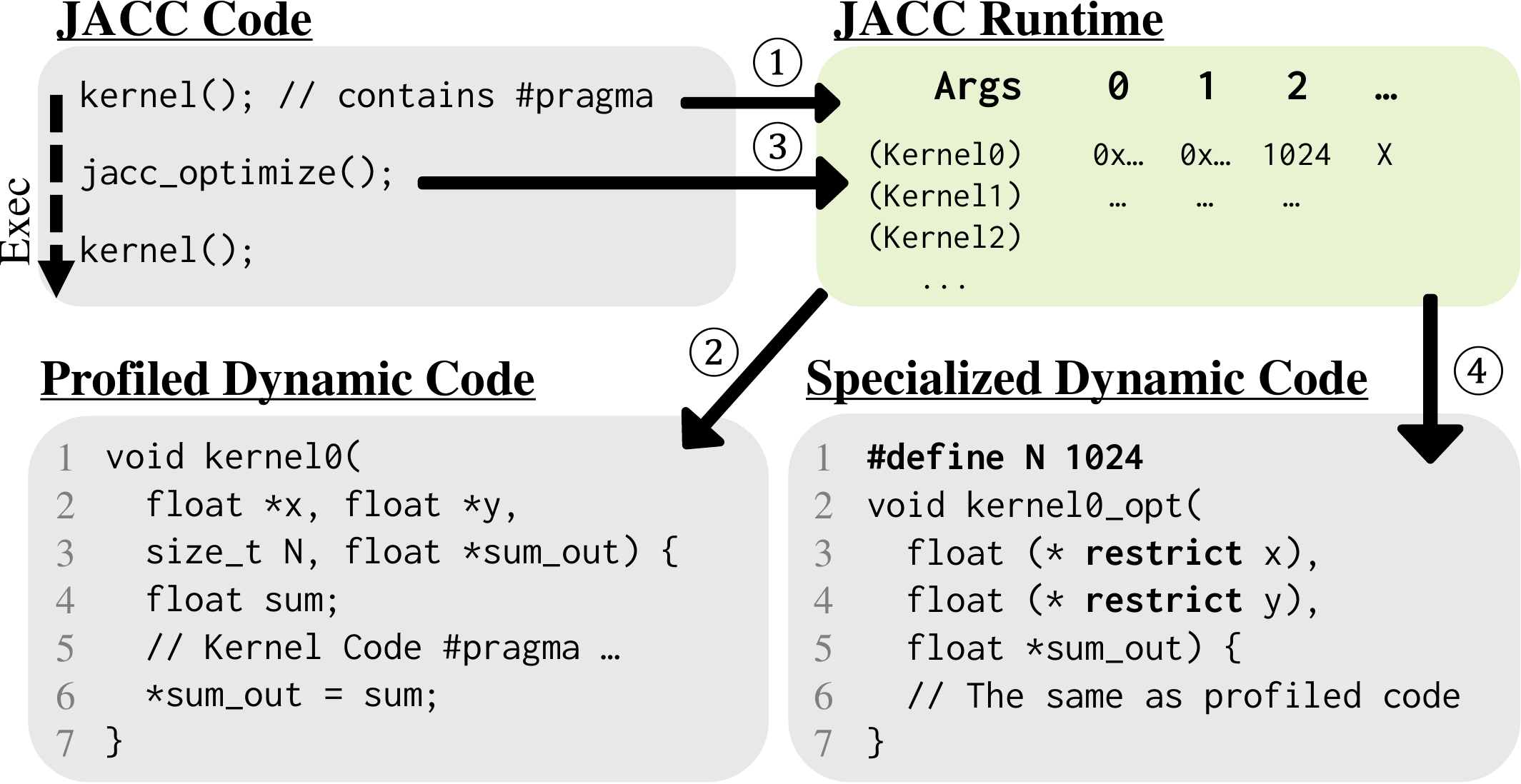}
  \vspace{-20pt}
  \caption{Compilation flow with on-the-fly kernel specialization. To collect runtime information about kernel arguments, mainly addresses of pointers and values of variables, profiling execution is conducted \circled{1} with original kernel declarations \circled{2}. With a user-invoked compilation event \circled{3}, specialized code is generated based on the profile results and thereafter used for succeeding kernel execution \circled{4}. The \texttt{sum} variable in the profiled code is exported through the \texttt{sum\_out} pointer, thus it is kept for specialized code as a dynamic variable (notated as "\texttt{X}" in the argument log of the JACC runtime).}
  \label{fig:spec}
  \vspace*{-0.08cm}
\end{figure}

We attempt to refine resource use by attaching runtime infor- mation to
function code so as to enable aggressive optimization in kernel compilation.
The compilation flow is shown in Fig.~\ref{fig:spec}.
For specialization, profile execution is conducted before the optimiza- tion;
then, at an additional compilation event (\texttt{jacc\_optimize} of JACC code in Fig.~\ref{fig:spec}),
parameters being invariable during
the execution are substituted with constants (Line 1 of specialized code in Fig.~\ref{fig:spec}) to
lower on-chip resource use.
Besides that,
pointer references that never conflict with others are declared with
the \texttt{restrict} keyword
to ensure intra-kernel parallelism (Lines 3-4 of specialized code in Fig.~\ref{fig:spec}). Even though user-invoked events can be substituted automatically by existing just-in-time compilation techniques~\cite{10.1145/2076021.2048126}, we explicitly invoke the desired compilation optimizations to measure the potential performance benefits.

\section{Multi-GPU Utilization with Predicates}\label{sec:multigpu}

\begin{figure}[t!]
\vspace*{6pt}
  \centering
  \begin{lstlisting}[backgroundcolor=\color{backgrey}, numbers=left, xleftmargin=3em, framexleftmargin=3em]
a[i]=x; b[i]=a[i]; x=c[j]; a[k]=x; b[k]=a[k];
\end{lstlisting}
  \begin{lstlisting}[backgroundcolor=\color{backblue}, numbers=left, xleftmargin=3em, framexleftmargin=3em]
/* a[i]=x */
((a_lb<=i && a_ub>=i)||
 (b_lb<=i && b_ub>=i)) ? a[i]=x:a[i];
/* b[i]=a[i] */
((b_lb<=i && b_ub>=i)) ? b[i]=a[i]:b[i];
/* x=c[j] */
x=((a_lb<=k && a_ub>=k)||
   (b_lb<=k && b_ub>=k)) ? c[j]:0;
/* a[k]=x */
((a_lb<=k && a_ub>=k)||
 (b_lb<=k && b_ub>=k)) ? a[k]=x:a[k];
/* b[k]=a[k] */
((b_lb<=k && b_ub>=k)) ? b[k]=a[k]:b[k];
\end{lstlisting}
\vspace*{-10pt}
\caption{Example of predicate-based filtering in C code. Original (top) and filtered code (bottom). References to array \texttt{a} have pred- icates for updating array \texttt{b} and itself (Lines 2-3 and 10-11), the references to array \texttt{b} have for itself (Lines 5 and 13), and the reference to array \texttt{c} has for array \texttt{a} and \texttt{b} (Lines 7-8).}
\label{fig:conv-code}
  \vspace*{0.5cm}
  \centering
  \hspace*{-0.15cm}
  \vspace*{0cm}
  \includegraphics[width=0.5\textwidth]{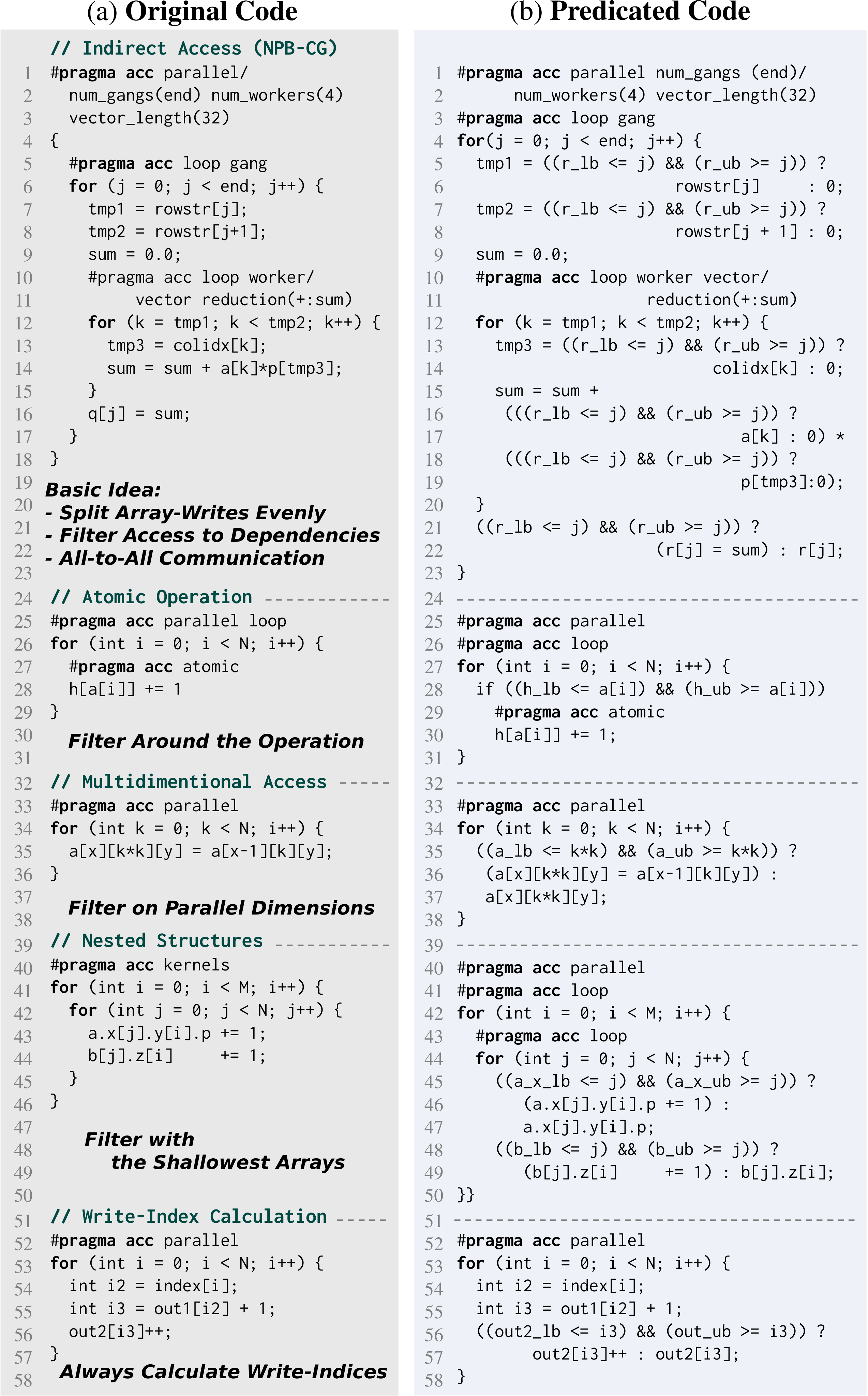}
  \vspace{-20pt}
  \caption{Original code (left) and corresponding predicated code (right)}
  \label{fig:sample}
  \vspace*{-0.4cm}
\end{figure}

We further improve utilization of intra-kernel parallelism %
by enabling
multi-GPU execution with JACC.
Whereas previous stud- ies have persistently focused on loop splitting over plural GPUs~\cite{10.1007/978-3-319-69953-0_7,komoda2013integrating},
this work divides data regions that each GPU updates to support real applications that usually entangle memory accesses among loop iterations.

\subsection{Predicate-Based Filtering}
Our technique, named {\it predicate-based filtering},
limits memory accesses depending on data regions that the GPU writes to,
assuming that redundant computational code and parameters do not degrade performance
due to low computational latency and high memory latency on GPUs.
First, we introduce data ranges for each updated array so that
array writes can be filtered based on the assigned range.
For instance, in C code, array write \ctext{\texttt{a[i]*=2}} is rewritten to 
\ctext{\texttt{(a\_lb <= i \&\& a\_ub >= i) ? a[i]*=2 : a[i]}}, where \texttt{a\_ub} and \texttt{a\_lb} indicate the upper and lower bound of array \texttt{a}, that are specified depending on the GPU. In Fortran, since there is no nested assignment, we use IF statement for filtering, with subsequent ELSE statement which contains an assignment of the same expression ( \ctext{\texttt{a(i)=a(i)}} ) that is later optimized away but facilitates compiler analysis.
Additionally,
we develop data-flow analysis for the innermost parallel region in each kernel
to detect data dependencies between arrays.
Then, we filter them to restrict accesses while solving dependencies as shown in Fig.~\ref{fig:conv-code}.
This analysis converts both array and variable references into the static single assignment (SSA) form, and iteratively finds dependencies among array accesses.

Updated data are sent to all other GPUs after each kernel execution to establish data coherency.
Device-memory allocations and host-to-GPU communications are replicated on all the GPUs and
the primary GPU is used for GPU-to-host transfers.
To guarantee the result of our analysis,
we check kernel arguments so as to duplicate computation and disable communications on
data that are referred through more than two pointers
which at least one of them is read and one is written (i.e. aliased pointers, which are usually avoided in OpenACC programs for loop independence).
When several pointers share the same array to update,
we merge their access ranges to follow the widest.
The necessary compu- tation for array-write indexing is always duplicated.
Regarding reduction or variable writes that are explicitly exported to host,
we filter the computation based on the range of the outermost parallel iterator.

Fig.~\ref{fig:sample} shows the actual cases of the transformation to predicated code.
Indirect accesses are supported just by filtering dependencies
which include loop ranges and array indices (Lines 5-8 and 13-14 in Fig.~\ref{fig:sample} (b)).
The OpenACC atomic operation is converted to be predicated around the operation (Lines 28-30 in Fig.~\ref{fig:sample} (b)).
We filter nested structures at the shallowest array accesses of the member references since otherwise their ranges could be vector values that require additional array accesses during kernel execution (Lines 45-49 in Fig.~\ref{fig:sample} (b)). %
The calculation of array-write indices is always outside the filters to be used in predicates (Lines 54-55 in Fig.~\ref{fig:sample} (b)).
The detail of multidimensional access is described in the following Section~\ref{subsec:multidim}.
Those are common patterns in OpenACC programs while precise data-dependency analysis such as polyhedral computation is hardly possible to treat the parallelism due to the complexity of the analysis and the limitation to affine loops~\cite{1639500,10.1145/2544100}.

\begin{figure}[b]
  \centering
  \begin{lstlisting}[caption={Kernel code from NPB-BT. Two inner loops are unrolled in actual code. Linear splits cause all-to-all dependencies among statements.}, label={jacc-sample4}]
#pragma acc parallel loop gang
for (i = 1; i <= gp02; i++) {
 #pragma acc loop worker vector
 for (k = 1; k <= gp22; k++) {
  for(m = 0; m < 5; m++)
   for(n = 0; n < 5; n++) {
    lhsY[n][m][BB][jsize][i][k] =
      lhsY[n][m][BB][jsize][i][k]
      - lhsY[n][0][AA][jsize][i][k]
      * lhsY[0][m][CC][jsize-1][i][k]
   /* - lhsY[n][1..3][AA][jsize][i][k]
      * lhsY[1..3][m][CC][jsize-1][i][k] */
      - lhsY[n][4][AA][jsize][i][k]
      * lhsY[4][m][CC][jsize-1][i][k];
    }}}
\end{lstlisting}
  \vspace*{-0.2cm}
\end{figure}

\subsection{Division of Multidimensional Arrays}
\label{subsec:multidim}

While being applicable to all OpenACC kernels as far as array writes are concerned, our filtering technique needs to duplicate execution on each GPU
when references between split ranges (such as all-to-all dependencies in Listing~\ref{jacc-sample4}) are found inside the kernel.
We alleviate this restriction by leveraging dimensional information.

If multidimensional arrays are linearly split regardless of the dimensional characteristic, the data dependency could be dispersed to the entire sections of array accesses. For example, the write to \texttt{lhsY} in Listing~\ref{jacc-sample4} (Lines 7-14) would be filtered for succeeding reads; thus, all the computations would be duplicated on each GPU. Here, we utilize parallel iterators (such as \texttt{i} and \texttt{k} in Listing~\ref{jacc-sample4}) to locate \textit{parallel dimensions}, where arrays can be split without duplicated computation. Based on the number of iterators each dimension contains,
we select the parallel dimension for each updated array to have the most parallel iterators while containing the least sequential iterators (such as \texttt{m} and \texttt{n}). When there are several candidates, we choose the leftmost dimension in the C language and the rightmost dimension in Fortran to gain better performance with suitable accesses to the memory layout (row-major and column-major order, respectively).

Each kernel execution is performed while equally dividing parallel dimensions among GPUs and accompanied by the GPU-to-GPU communication through \texttt{cudaMemcpy2DAsync}. Each array is concurrently transferred regarding other data and other GPUs. We synchronize GPUs at the beginning and ending of the communication.

\subsection{Adaptive Utilization}

In order to avoid lower performance due to data distribution overheads,
we enable multi-GPU execution for each kernel in an adaptive way,
while otherwise duplicating computation on all GPUs and performing no GPU-to-GPU communication.

First, we start the execution on the mode of duplication. After an initial warm-up run, we profile the average ratio of array-write size ($\mathrm{WriteSize}$) to execution time ($time_{\mathrm{Kernel}}$) as $eff_{\mathrm{dup}}$, until we observe five executions that satisfy the requirement to have the peak performance be better than duplication:
\vspace{0.5em}
\begin{equation}
  time_{\mathrm{Kernel}}\ \ >\ \ time_{\mathrm{Kernel}} / n_{\mathrm{GPUs}} + \mathrm{WriteSize} / peak_{\mathrm{P2P}}
\end{equation}
Here, $peak_{\mathrm{P2P}}$ is the unidirectional bandwidth of one GPU-to-GPU connection (e.g. $25\mathrm{GB/s}$ in NVIDIA DGX-1) and $n_{\mathrm{GPUs}}$ is the number of GPUs used.

After switching to multi-GPU execution,
we disable it when either one of the two following conditions is satisfied at least five times and the average difference of the left value and the smaller right value goes above zero in equations (2-3).
\begin{enumerate}[topsep=0.8em, leftmargin=0.2in,rightmargin=0.05in, itemsep=8pt]
\item The total execution time including the communication time ($time_{\mathrm{Comm}}$) becomes longer than the kernel execution time multiplied by $n_{\mathrm{GPUs}}$:

\vspace{-.5em}
\begin{equation}
  time_{\mathrm{Kernel}} + time_{\mathrm{Comm}}\ \ >\ \ time_{\mathrm{Kernel}} \times n_{\mathrm{GPUs}} \ 
\end{equation}

\item The total execution time surpasses the profiled execution time of duplication:

\vspace{-.5em}
\begin{equation}
  time_{\mathrm{Kernel}} + time_{\mathrm{Comm}}\ \ >\ \ eff_{\mathrm{dup}} \times \mathrm{WriteSize}
\end{equation}
\end{enumerate}
The first condition excludes the case that the GPU-to-GPU commu- nication has larger latencies than expected. The second prevents performance degradation caused by kernels that are unsuccessfully parallelized.

\subsection{Implementation}
\begin{figure}[t]
  \vspace{5pt}
  \centering
  \includegraphics[width=0.49\textwidth]{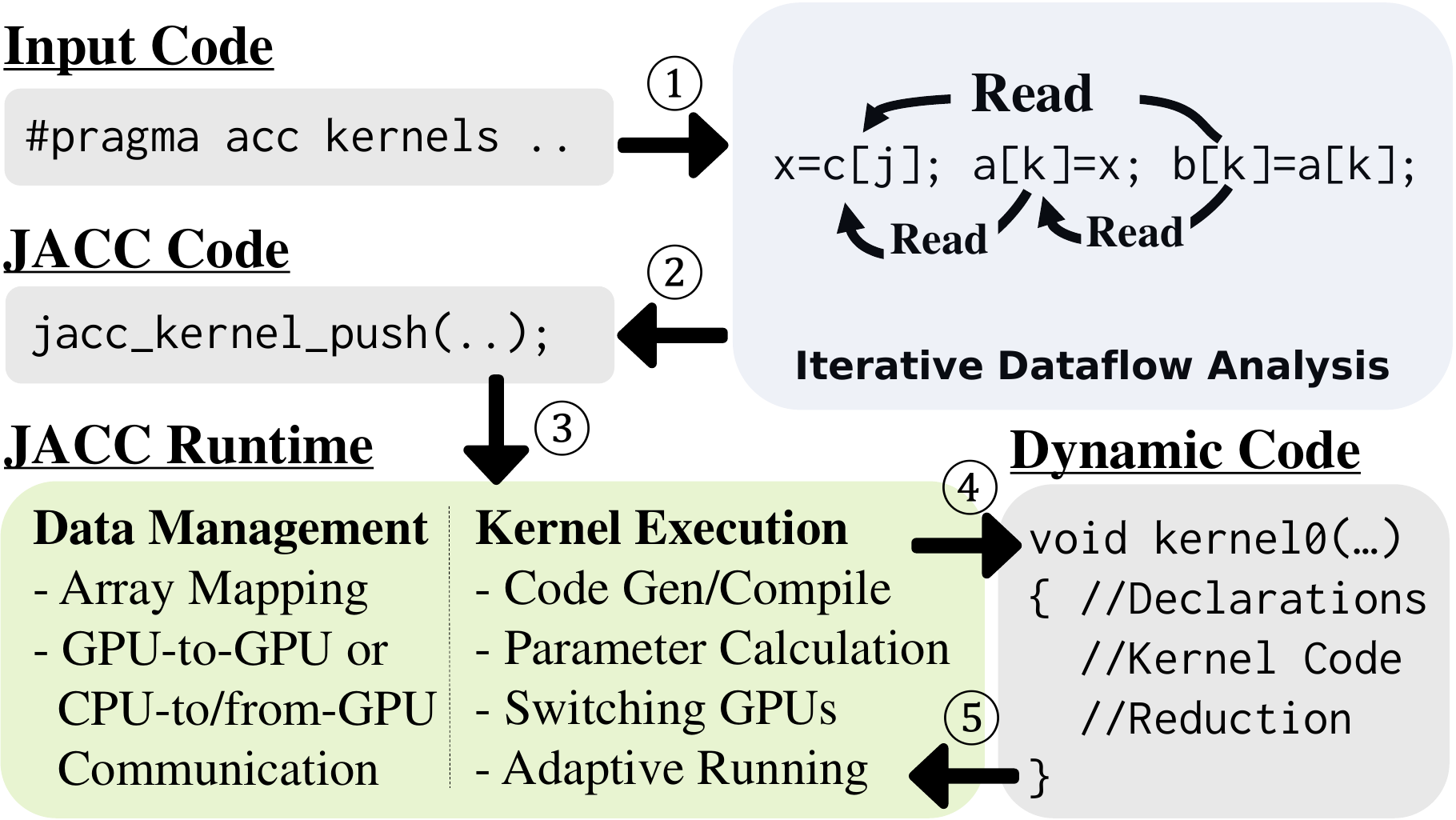}
  \vspace{-20pt}
  \caption{Execution flow of predicate-based filtering. Having the results of iterative dataflow analysis \circled{1}, JACC code is generated from the input \circled{2}. On the execution \circled{3}, dynamic code is created from kernel code, extended declarations and reduction code in accordance with runtime information \circled{4}. Once the dynamic code is compiled with a specified compiler, the kernel execution is conducted with dynamic parameters while switching GPUs and reflecting results to the host and each other devices \circled{5}.}
  \label{fig:jacc-exec}
\end{figure}

We integrate predicated-based filtering into JACC, which trans- lator is implemented as a XcodeML~\cite{xcodeml} converter. The execution flow of predicate\-/based filtering is shown in Fig.~\ref{fig:jacc-exec}. From OpenACC code in C or Fortran, our implementation generates JACC code, in which kernel code is embedded as strings. Although the kernel code can be translated at runtime, we apply predicated-based filtering beforehand for our experiments; thus, the embedded kernel code already has the predicates. JACC's runtime code generator is utilized for setting array ranges and constructing multi-GPU reduction code based on the arguments of runtime routines.
Runtime overheads of JIT dynamic compilation are well known, and itself a target of research~\cite{10.1145/2076021.2048126}, however it is outside the scope of this work and left as a possible future area for optimization. Thus, we evaluate the performance without dynamic compilation overheads (Section~\ref{subsubsec:kernel-spec}), which is on average about 2 seconds per kernel for the initial compilation.

For simultaneous execution of multiple GPUs, we use OpenMP rather than OpenACC's asynchronous mechanism that holds some non-negligible latencies~\cite{6495877, 10.1007/978-3-319-74896-2_9}. OpenMP's pragmas are put only inside JACC's library. Whereas GCC does not allow the mix of OpenACC and OpenMP, our separated-compilation strategy re- alizes a combinatory use for both PGI and GCC. The OpenMP use is not inherent here and our techniques introduced in this paper are general enough to support other compilers and other programming models.%

\begin{figure}[t]
  \vspace{5pt}
  \centering
  \includegraphics[width=0.475\textwidth]{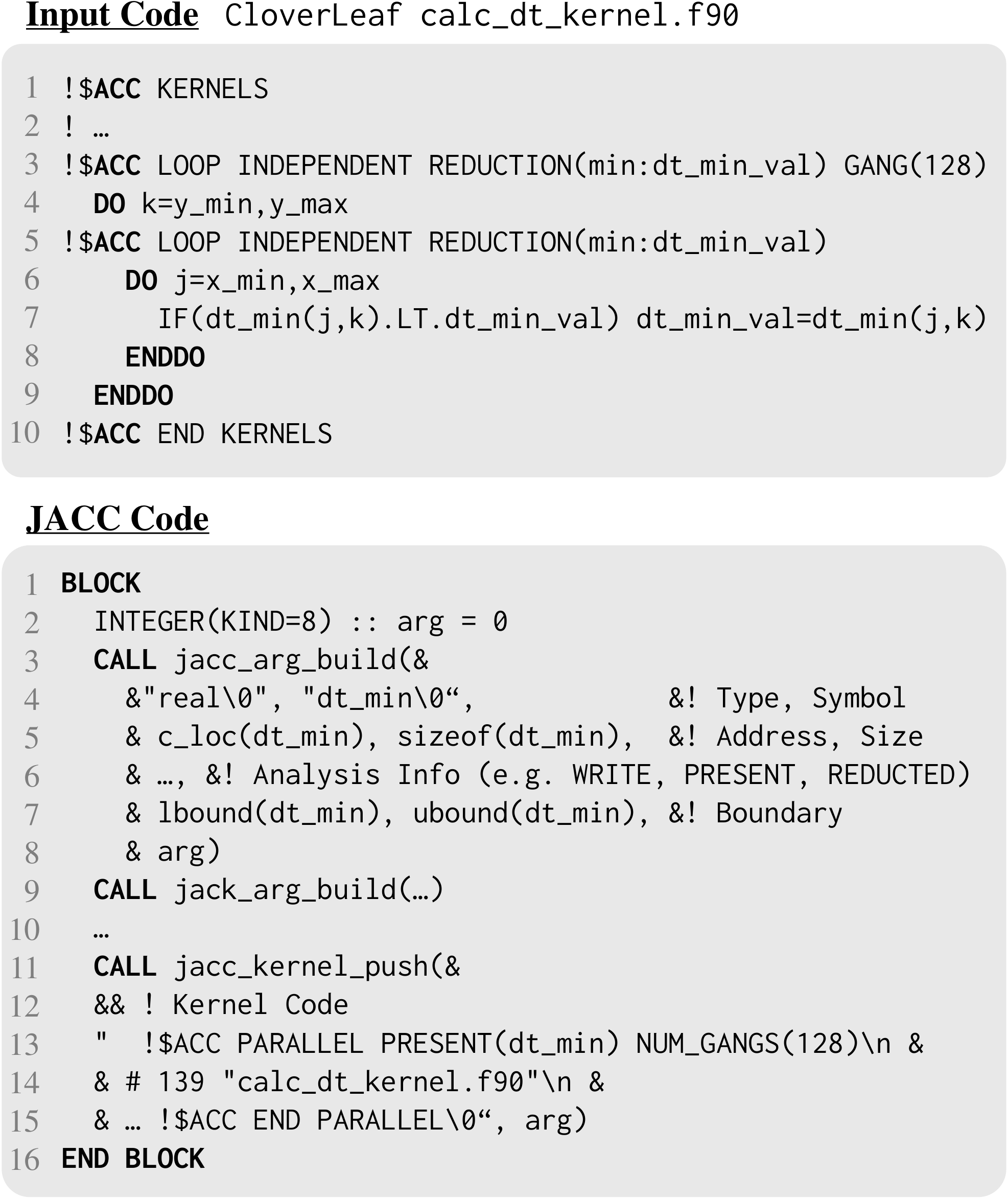}
  \vspace*{-20pt}
  \caption{Actual case of JACC code generation in Fortran from the input (top). The arguments and kernel code are partially shown in JACC code (bottom).}
  \label{fig:fortran}
  \vspace*{-0.2cm}
\end{figure}

Fig.~\ref{fig:fortran} shows actual JACC code in Fortran.
The kernel arguments are built with static and runtime parameters to create dynamic code correctly
and the kernel code is set with line information for debugging purposes (Lines 3-8 and Line 14 in JACC code of Fig.~\ref{fig:fortran}, respectively).
For predicate-based filtering,
JACC's runtime automatically launches the dynamic code
on each GPU and performs GPU-to-GPU communications
based on those passed arguments.

\section{Methodology} \label{sec:method}

\subsection{Hardware and Software}

We measure the performance changes of our proposed techniques using the NVIDIA Tesla V100 SXM2 GPUs (16GB Memory) on NVIDIA DGX-1.
DGX-1 contains eight GPUs, where each four GPUs are interconnected with NVLink in an all-to-all fashion. Additionally, each GPU has one NVLink connection to one of the other four GPUs, while the remaining GPUs and CPUs are connected with PCI Express Gen3 x16.
We use only the tightly coupled four GPUs to perform our experiments where each link offers a unidirectional bandwidth of 25GB/s while $\mathrm{GPU_0}\leftrightarrow\mathrm{GPU_3}$ and $\mathrm{GPU_1}\leftrightarrow\mathrm{GPU_2}$ are dually linked. Tesla V100 has a peak single-precision performance of 15.7 TFLOP/s and a peak memory throughput of 900GB/s.
DGX-1 uses dual 20-core Intel Xeon processors.

For the compilation, we use the PGI compiler 20.9 and GCC 10.2.0 with CUDA 11.0.
We compile OpenACC kernels and our runtime library using the following sets of compiler arguments:
\ctext{\texttt{"-O2 -acc -mp -ta=tesla:cc70 -Mcuda"}} for PGI and 
\ctext{\texttt{"-O2 -f(openacc|openmp) -foffload=nvptx-none -foffload=-lm -fno-strict-aliasing"}} for GCC.
Currently, the Fortran trans- lation is tested only for predicate-based filtering with PGI.
The experiments with GCC are conducted while omitting the \texttt{worker} parallelism (described in Section~\ref{sec:background}) so as not to exceed the maximum number of threads allowed in GCC per thread-block.

\subsection{Benchmarks}

\begin{table*}[b]
  \centering
  \caption{Benchmark description}
  \label{tab:bench}
\vspace{-0.4cm}
\def\arraystretch{1}
\setlength\tabcolsep{9.1pt}
\setlength{\extrarowheight}{1pt}
\begin{tcolorbox}[colback=white,boxsep=0pt,left=2pt,right=2pt,top=1pt,bottom=1pt, arc=0mm, boxrule=\arrayrulewidth, colframe=black, enlarge top by=-0.1cm, enlarge bottom by=-0.1cm]
\footnotesize
  \begin{tabular}{c|l|c|c|r|c}
    Name & Description & Dependency & Problem Size & Memory\ \ & Num Kernels \\
\hline
\rowcolor{lightgrey}
    BT & CFD with Block Tri-Diagonal Solver
    & Halo (3D) & Class C: 162x162x162 (FP64), iter=200 ~  & 4.915 GB & 46
    \\
\hdashline[1pt/2pt]
    CG & Minimum-Eigenvalue Calculation %
    & Irregular & Class C: size=150000 (FP64), iter=3750 ~ & 0.747 GB & 16\\
\hdashline[1pt/2pt]
\rowcolor{lightgrey}
    EP & Random-Number Generation in Parallel %
    & None & Class D: size=137438953472 (FP64)~ ~~~~ & 2.305 GB & 4\\
\hdashline[1pt/2pt]
    FT & Discrete 3D Fast Fourier Transform
    & All-to-All  &  Class C: 512x512x512 (FP64), iter=20~ ~~  & 9.317 GB & 12\\
\hdashline[1pt/2pt]
\rowcolor{lightgrey}
    LU & CFD with Lower-Upper Gauss-Seidel Solver
    & Halo (3D)  & Class C: 162x162x162 (FP64), iter=250 ~ & 1.471 GB & 59\\
\hdashline[1pt/2pt]
    MG & Multigrid Discrete Poisson Equation
    & Long \& Short & Class C: 512x512x512 (FP64), iter=1000~ & 6.114 GB & 16\\
\hdashline[1pt/2pt]
\rowcolor{lightgrey}
    SP & CFD with Scalar Penta-Diagonal Solver
    & Halo (3D) & Class C: 162x162x162 (FP64), iter=400 ~  & 1.700 GB & 65\\
\hline
   CloverLeaf & 2D Euler Equations Solver & Halo (2D) & 3840x1920 (FP64), step=1800 & 1.749 GB & 114 \\
\hdashline[1pt/2pt]
\rowcolor{lightgrey}
   CCS-QCD & Lattice QCD Simulation & Halo (3D) & 32x32x32x128 (FP64), iter=1000 for BiCGStab & 15.255 GB & 27 \\
\hdashline[1pt/2pt]
   Himeno  & 19-point Jacobian Stencil Computation  & Halo (3D) & Size XL:1024x512x512 (FP32), iter=1000 & 14.409 GB & 2
  \end{tabular}
\end{tcolorbox}  
\end{table*}

For the evaluation,
we use a manually-tuned OpenACC version of NAS Parallel Benchmarks (NPB) written in C~\cite{xu2014parallel} and three Fortran mini-apps:\hspace{0.1em}CloverLeaf~\cite{cloverleaf},\hspace{0.1em}CCS-QCD~\cite{ccs-qcd} and the Himeno benchmark~\cite{himeno}.
Each benchmark of NPB, briefly described in Table~\ref{tab:bench},
is executed with the largest problem size for the target GPUs (Class C)
except EP, where we choose Class D for longer execution.
Moreover, to prolong the execution time of CG and MG,
we multiply the number of iterations by 50.
With regard to CloverLeaf, we select the same input as the SPEC ACCEL benchmark suite~\cite{spec},
while maximizing the GPU memory utilization of the other two mini-apps.

EP is the only application to be compute-bound for arithmetic operations of random-number generation involving fewer array accesses, while other benchmarks become memory-bound on the state-of-the-art accelerators. There is no data dependency among parallel threads in EP.

BT/LU/SP deal with three-dimensional computational fluid dynamics (CFD) with different solvers.
BT updates multidi- mensional arrays from 3D to 6D, especially 4D to 6D have the leftmost 1D to 3D dimensions for loop-independent indices as shown in Fig.~\ref{jacc-sample4} (Lines 7-14), respectively.
LU/SP use 3D to 4D arrays in a similar fashion to BT.
LU has relatively quicker execution for each kernel compared to BT/SP and
the latency of BT mainly consists of the execution of several time-consuming kernels.

FT conducts 3D fast Fourier transform (FFT) incurring all-to-all accesses, and
MG is the benchmark of multigrid computation which require long and short communications.
CG calculates a minimum eigenvalue of a sparse symmetric positive matrix with the conjugate gradient method, causing irregular accesses to an updated 1D array.

CloverLeaf is a hydrodynamics mini-application consisting of 100+ kernels which sufficiently demonstrates the workflow of real-world applications.
Based on Euler's method, 2D stencil grids are updated by each kernel having minimum control logics and halo accesses through double buffers for avoiding dependencies among loop iterations.

CCS-QCD simulates lattice quantum chromodynamics (QCD) with a linear-equation solver for a large sparse matrix in 3D. The execution is mainly composed of the biconjugate gradient stabilized method (BiCGStab) along with neighboring transfers and all-to-one reduction.
The reduction accumulates slight errors upon each execution which affect the total number of iterations
when the computational order is changed due to multi-GPUs.
Therefore, we fix the number of iterations for BiCGStab to 1,000.

Himeno iteratively updates a 19-point stencil grid according to Jacobi's method. The code structure is far simpler than the other two mini-apps. The kernels are constructed from three nested loops where each iteration corresponds to the grid's dimension and no dependence exists among them. Optimally, 3D halo accesses solve inter-kernel dependencies, and computational errors are reduced after kernel execution.

We report performance after
an initial warm-up run that causes runtime compilation and profiling for adaptive utilization.
The benchmark-reported data is quoted for the result of NPB and the total execution times for the Fortran mini-applications.

\section{Results}\label{sec:results}

\begin{figure}[b]
  \centering
  \vspace*{0.7cm}
  \includegraphics[width=0.49\textwidth]{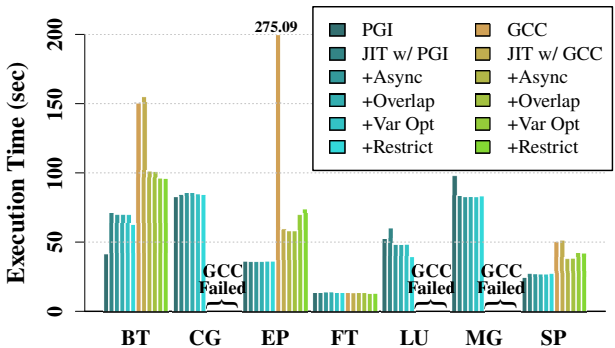}
  \vspace{-25pt}
  \caption{Async \& kernel opt on NVIDIA Tesla V100 SXM2}
  \label{fig:v100}
  \vspace*{0.5cm}
\end{figure}

\subsection{Basic Extension}
\label{subsec:basic-ext}

Fig.~\ref{fig:v100} shows the performance changes with JACC's basic extension
for both the PGI compiler and GCC using NPB.
The \ctext{\texttt{JIT w/ (PGI/GCC)}} bars indicate the performance of converted code without any optimization. Here, only one asynchronous queue is used with \ctext{\texttt{+Async}}, whereas 16 queues are used with \ctext{\texttt{+Overlap}}. Along with that, the \ctext{\texttt{+Var Opt}} execution adds kernel optimization with constant parameters transformation discussed in Section~\ref{subsubsec:kernel-spec}. Furthermore, \ctext{\texttt{+Restrict}} adds \texttt{restrict} to pointers.
Since GCC produces incorrect results with the original code of CG/LU/MG,
they are omitted from results.

First, performance degradation is observed for converted code compared to original code in the case of BT/LU with PGI, where generated code fails to leverage static array sizes for some optimization at compile time because static arrays are separately declared.
However, improvements are observed in the case of MG with PGI and EP with GCC.
Otherwise, original performances are mostly kept.
On average, asynchronous execution with single queue achieves better performance
by 3.43\% with PGI and 22.08\% with GCC, respectively.
However, the time-consuming kernels in each benchmark prevent overlap execution; thus,
\ctext{\texttt{+Overlap}} does not improve the performance from \ctext{\texttt{+Async}}.
With \ctext{\texttt{+Restrict}}, we achieve better performance up to 23.39\%
in the case of BT/LU with PGI and 5.59\% less performance in EP with GCC.
The performance difference between PGI and GCC is primarily caused by the latency of memory allocation; PGI owns memory pools for device memory, while GCC does not.

The \ctext{\texttt{+Var Opt}} version has no performance change in most cases and
rather worsen efficiencies in the case of EP/SP with GCC.
Further exploration showed that
some cases of \ctext{\texttt{+Var Opt}} suffer from limited arithmetic unit utilization caused by ineffective threads which are created due to reduced register use.
On the other hand, performance improvements of \ctext{\texttt{+Restrict}} are achieved by parallelized memory accesses which require additional registers.

\begin{figure*}[t]
  \centering
  \subfloat{\includegraphics[width=0.50\textwidth]{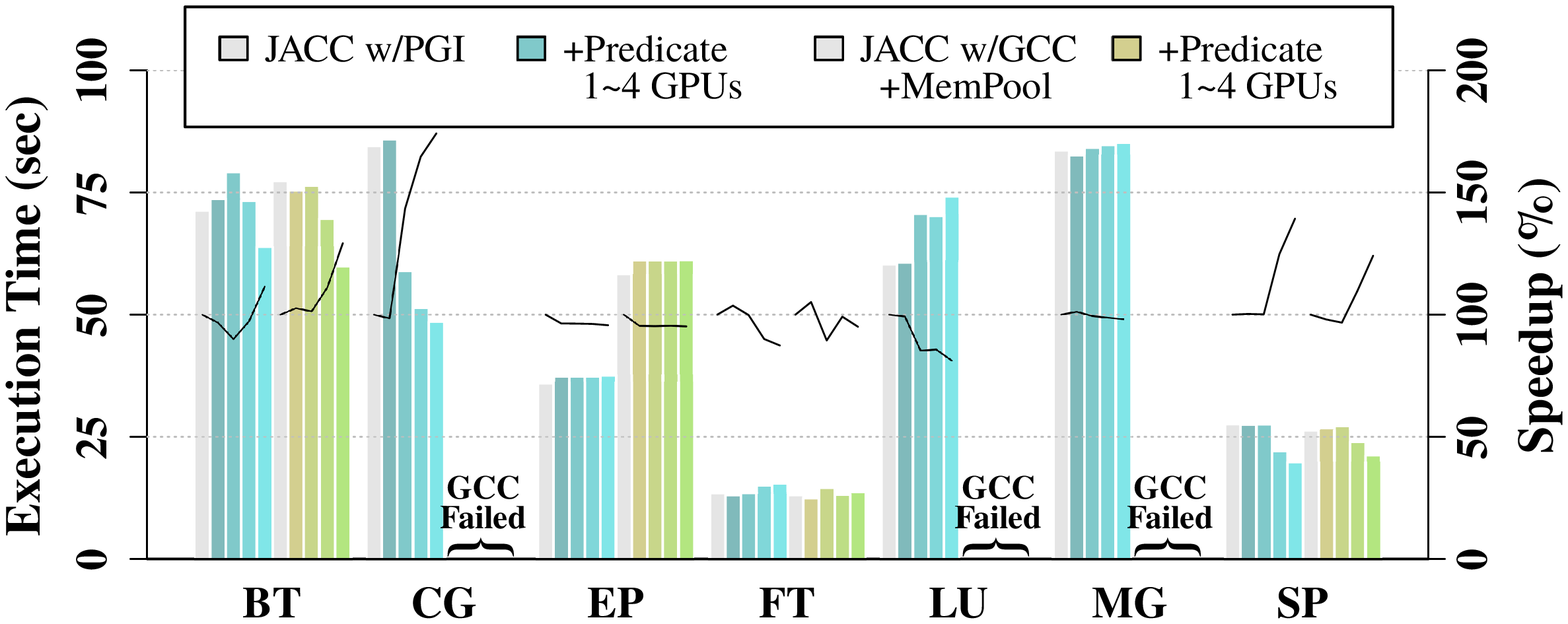}}
  \subfloat{\includegraphics[width=0.50\textwidth]{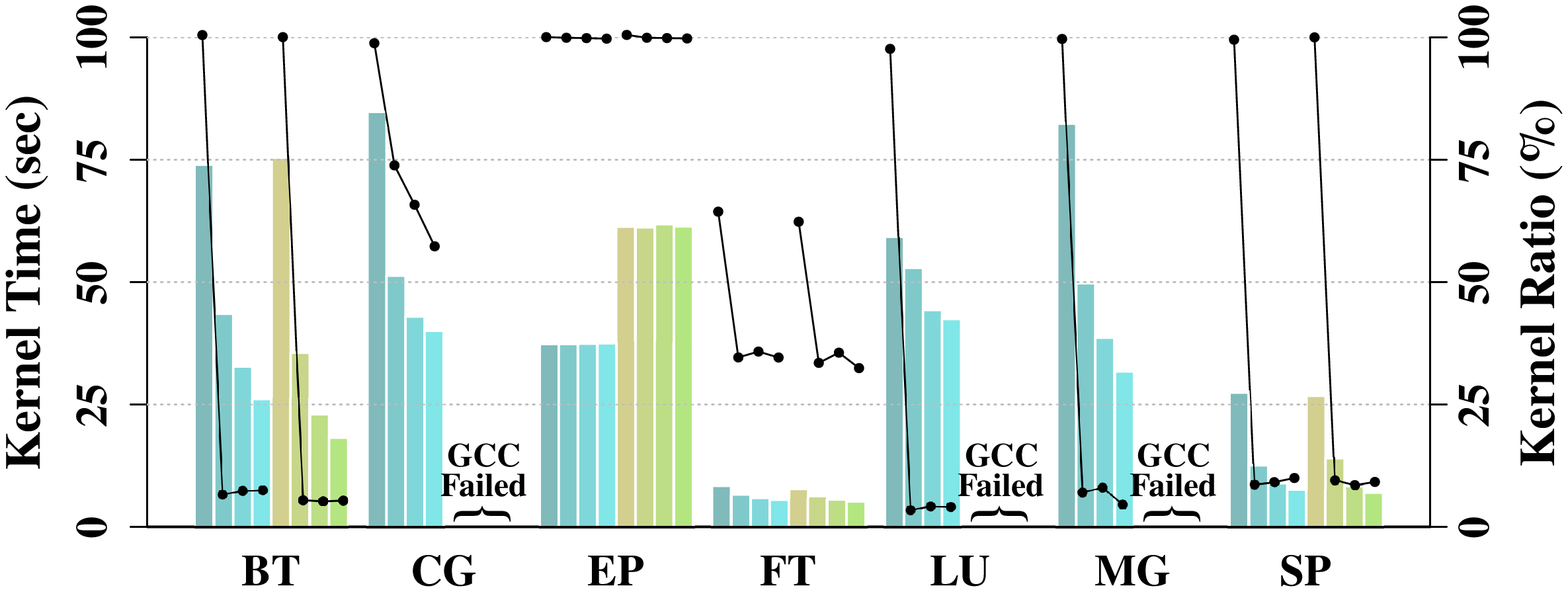}}\\
  \centering
  \vspace{-0.4cm}
  \subfloat{\includegraphics[width=0.50\textwidth]{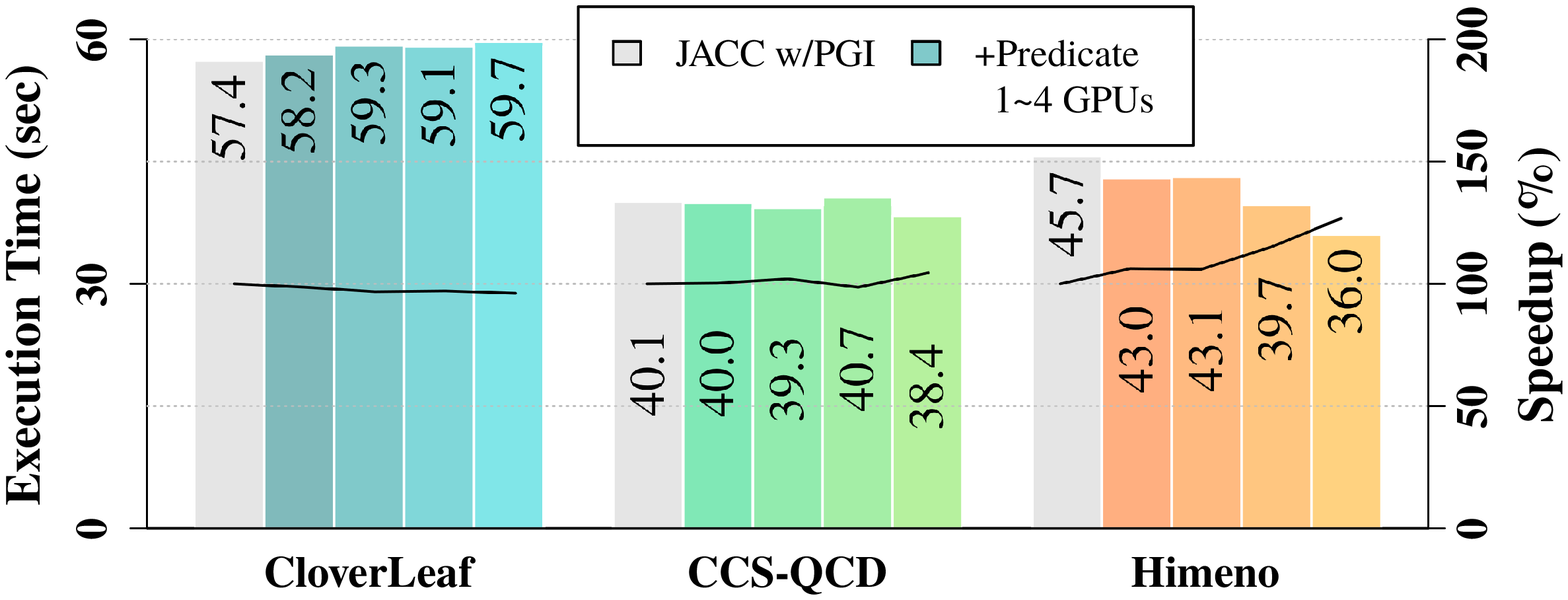}}
  \subfloat{\includegraphics[width=0.50\textwidth]{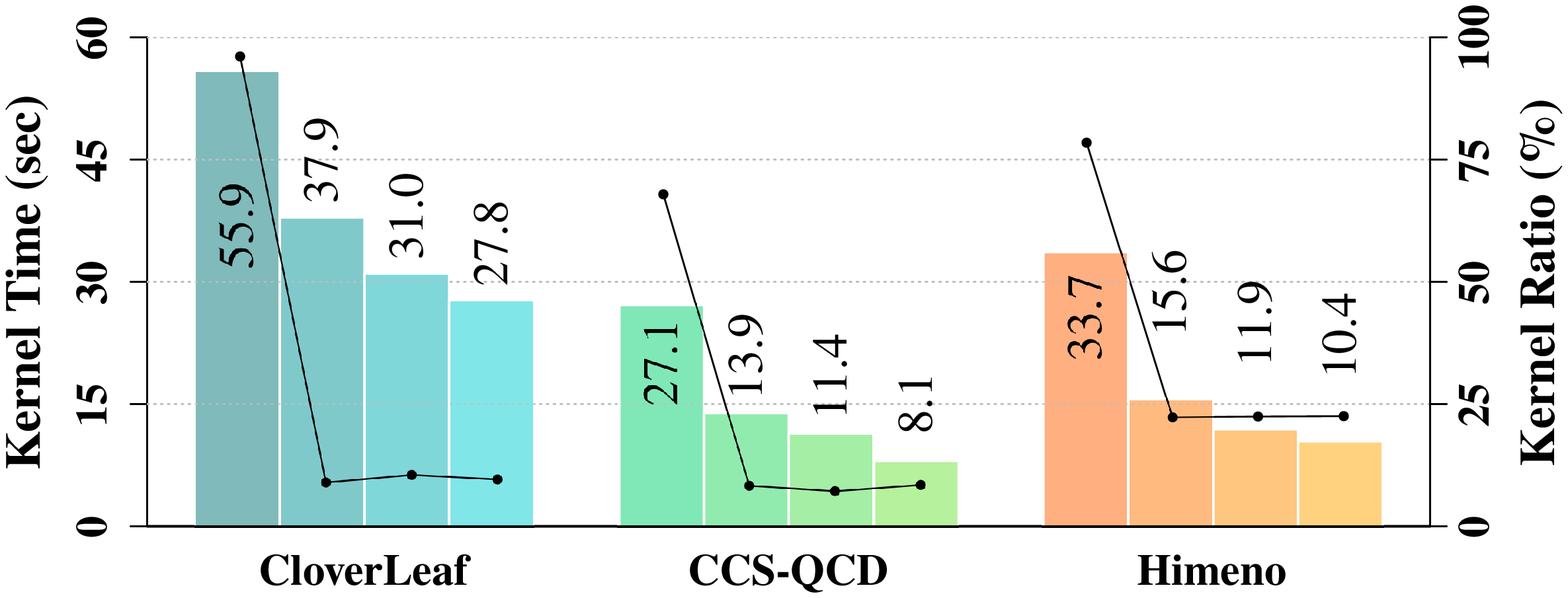}}\\
  \centering
  \vspace{-10pt}
  \caption{Performance scaling of predicate-based filtering using NPB with PGI/GCC (top) and using the Fortran mini-apps with PGI (bottom). Left figures provide the execution times with bars and the speedups with lines compared to the plain JACC code. Right figures show the kernel time when we enable multi-GPU execution with no adaptive algorithm; the kernel ratio to the total time is given by the line.}
  \label{fig:v100-2}
  \vspace{-0.2cm}
\end{figure*}

\begin{table*}[t]
  \centering
  \caption{\small Performance details with PGI in use of four GPUs (The result of duplicated execution on all the GPUs is used for the Kernel Dup column only; Other columns use the results of adaptive execution. The Kernel Adapted column shows the kernel execution time for multi-GPU adapted kernels with the average. The {\bf bold} values indicate performance improvements from the duplicated execution)}
  \label{tab:detail}
  \vspace{-0.4cm}
\def\arraystretch{1}
\setlength\tabcolsep{8.15pt}
\setlength{\extrarowheight}{1pt}
\hspace*{-0.4em}
\begin{tcolorbox}[colback=white,boxsep=0pt,left=2pt,right=2pt,top=1pt,bottom=1pt, arc=0mm, boxrule=\arrayrulewidth, colframe=black, enlarge top by=-0.1cm, enlarge bottom by=-0.1cm]
\footnotesize
  \begin{tabular}{c||c|c||c|c|r|r|r|r|r}
    \multirowcell{2}{Name} &
    \multirowcell{2}{Num\\ Kernels} &
    \multirowcell{2}{Kernel\\ Dup [ms]} &
    \multirowcell{2}{Num\\ Adapted} &
    \multirowcell{2}{Comm + \\ Kernel [ms]} &
    \multirowcell{2}{ \ \ Kernel Total \ \ \\ (ave) [ms]} &
    \multirowcell{2}{Kernel Adapted\\(ave)  [ms]} &
    \multirowcell{2}{Comm\\ (ave) [ms]} &
    \multirowcell{2}{Average\\ WriteSize} &
    \multirowcell{2}{GPU-to-GPU\\ Bandwidth}
    \\
    \multicolumn{1}{c||}{} &
    \multicolumn{1}{c|}{} &
    \multicolumn{1}{c||}{} &
    \multicolumn{1}{c|}{} &
    \multicolumn{1}{c|}{} &
    \multicolumn{1}{c|}{} &
    \multicolumn{1}{c|}{} &
    \multicolumn{1}{c|}{} &
    \multicolumn{1}{c|}{}
    \\
\hline
\rowcolor{lightgrey}
    BT & 46 & 88,715 & 3 & {\bf 63,856} & 46,639 ~~(0.44) & 14,940~~(24.75) & 17,217 ~(28.52) & 684.10 MB & 23.99 GB/s
    \\
\hdashline[1pt/2pt]
    CG & 16 & 75,178 & 7 & {\bf 44,137} & 38,720 ~~(0.08) & 34,365 ~~(0.12) & ~5,417 ~~(0.02) & ~~~0.10 MB & 5.40 GB/s
    \\
\hdashline[1pt/2pt]
\rowcolor{lightgrey}
    EP & 4 & 37,485 & 3 & 37,787 & 37,710~~(24.55) & 37,710~~(24.55) & ~~~~~77 ~~(0.05) &  ~~~0.03 MB & 0.54 GB/s
    \\
\hdashline[1pt/2pt]
    FT & 12 & ~8,472 & 4 & ~8,757 & ~6,096~~(47.29) & ~4,983~~(62.37) & ~2,661~~(33.31) & 806.92 MB & 24.23 GB/s
    \\
\hdashline[1pt/2pt]
\rowcolor{lightgrey}
    LU & 59 & 76,325 & 7 & {\bf 71,614} & 64,342 ~~(0.09) & ~3,886 ~~(0.03) & ~7,272 ~~(0.06) &  ~~~0.39 MB & 6.28 GB/s
    \\
\hdashline[1pt/2pt]
    MG & 16 & 83,586 & 3 & 83,654 & 83,654 ~~(0.47) & ~5,604 ~~(0.35) & ~~~~~~0 ~~(0.00) &  ~~~0.00 MB & 0.00 GB/s
    \\
\hdashline[1pt/2pt]
\rowcolor{lightgrey}
    SP & 65 & 27,809 & 3 & {\bf 19,609} & 16,648 ~~(0.64) & ~1,969 ~~(1.64) & ~2,961 ~~(2.47) & ~~58.24 MB & 23.60 GB/s
    \\
\hline
   CloverLeaf & 114 & 55,017 & 3 & 55,272 & 54,079 ~~(0.22) & 10,980 ~~(4.27) & ~1,193 ~~(0.46) &  ~~~5.46 MB & 11.76 GB/s
   \\
\hdashline[1pt/2pt]
\rowcolor{lightgrey}
   CCS-QCD & 27 & 26,517 & 11 & {\bf 24,641} & 13,031 ~~(0.97) & ~5,811 ~~(1.91) & 11,610 ~~(3.82) & ~~91.67 MB & 24.02 GB/s
   \\
\hdashline[1pt/2pt]
   Himeno & 2 & 33,726 & 1 & {\bf 23,149} & 11,894 ~~(5.93) & ~8,318 ~~(8.30) & 11,255~~(11.23) & 271.47 MB & 24.18 GB/s
   \\
  \end{tabular}
\end{tcolorbox}  
  \vspace{0.2cm}
\end{table*}

\subsection{GCC Custom Allocation}
\label{subsec:gcc}

 Since the original version of GCC suffers the performance degradation by GPU-memory allocation, we integrate memory pools into GCC's runtime library libgomp for our multi-GPU experiments to show explicitly the performance improvements by kernel parallelization. We prepare two pools: one is for user\-/invoked memory allocation such as through pragmas and runtime routines. Another is for runtime-managed allocation of variables and stacks which tends to be much smaller than the former. We manage those pools to keep unused memory segments and reuse them for new allocation by selecting the smallest but capable one on the device.

With the memory-pool integration,
GCC's efficiency becomes competitive to PGI while having -7.83\% $\sim$ 5.05\% better through- puts for the plain JACC code except the case of EP, where the kernel execution poses a 38.31\% overhead due to GCC's device-code efficiency, as shown in Fig.~\ref{fig:v100-2}.

\subsection{Multi-GPU Utilization}
\label{subsec:result-multi}
\subsubsection{Total Improvements \& Kernel Speedups}

We show the overall performance and kernel speedups with predicate-based filtering in Fig.~\ref{fig:v100-2}.
When compared to single-GPU execution,
the total execution time with our proposed technique is better in five among 10 evaluated benchmarks, from 4.05\% up to 43.43\% when enabling four GPUs.
Especially, when only the kernel execution and GPU-to-GPU transfers are concerned, six bench- marks (BT/CG/LU/SP/CCS-QCD/Himeno) improve the execution time by 23.9\% on average as shown in Table~\ref{tab:detail}. Other benchmarks still remain unchanged with some slight degradation up to 3.36\% while having several kernels enabled for multi-GPU execution. The noticeable slowdowns we observe in the total execution time of LU/MG are caused due to other factors necessary for multi-GPUs such as memory allocation and synchronization.
As an opposite fashion to \ctext{\texttt{+Var Opt}},
the predicate-appended code mostly holds the performance of the plain JACC code with single-GPU use. %

Profiling the kernel speedups with no adaptive execution showed that predicate-based filtering parallelizes many kernels.
Using the memory-intensive benchmarks BT/SP, PGI achieves 2.83x and 3.59x improvements on four GPUs and GCC does 4.13x and 3.85x, respectively.
For LU, however, the shorter than 1ms running time of each kernel execution limits acceleration to 1.40x, involving overheads for duplicating program structures on all the GPUs.
EP does not have any improvement due to its compute\-/bound nature.
Comparing adaptive and non-adaptive execution,
CG has almost the same improvement despite other benchmarks are prevented from full parallelization due to the high communi- cation\-/kernel ratios. For example, in Table~\ref{tab:detail}, around the 20\% execution of CloverLeaf is distributed over multi-GPUs but those kernels are not well enhanced, while the remaining execution is duplicated because of the excessive communication latencies, hence, resulting in no speedup.

\subsubsection{Data-Size Scaling}
\begin{figure*}[t]
  \centering
  \vspace{-0.4cm}
  \subfloat{\includegraphics[width=0.335\textwidth]{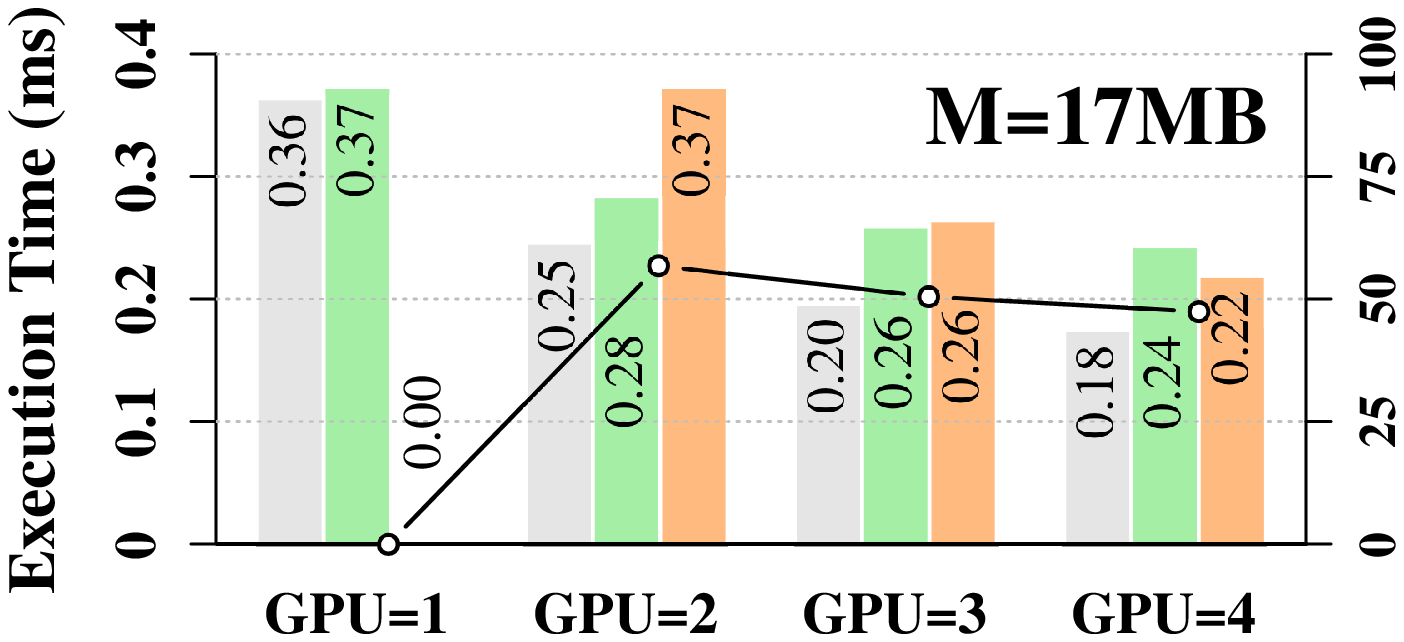}}
  \hspace{-0.36cm}
  \subfloat{\includegraphics[width=0.335\textwidth]{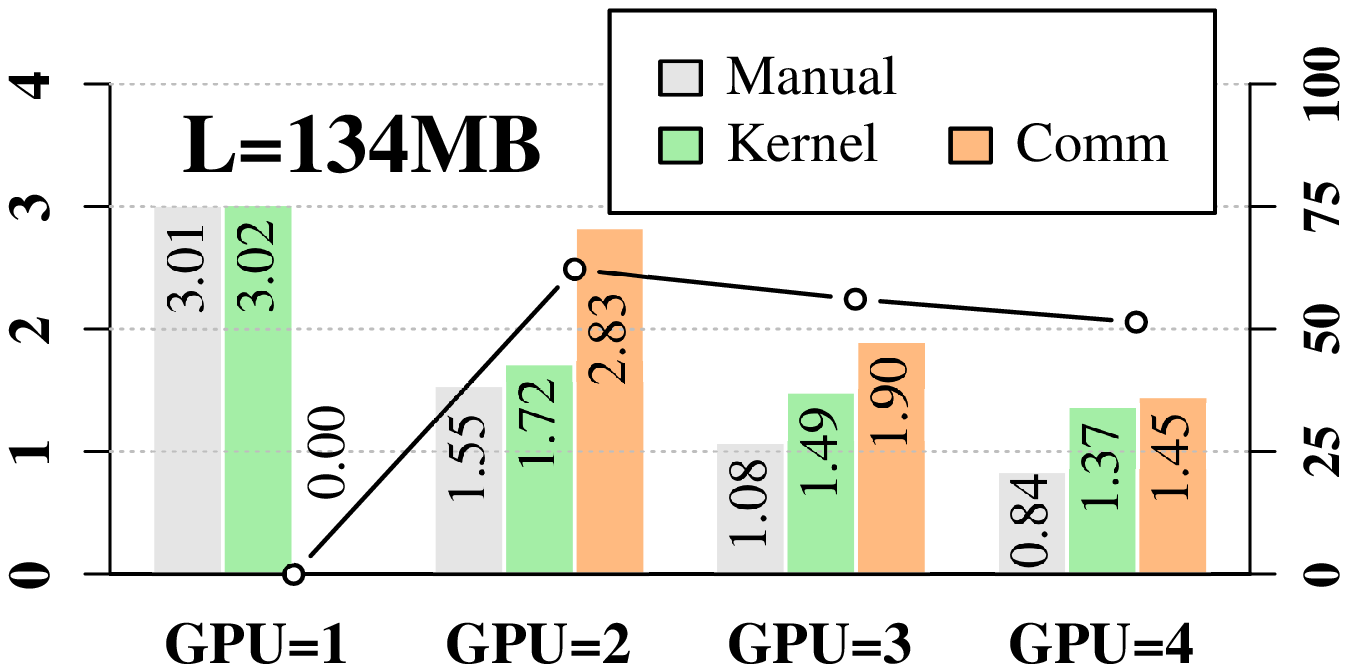}}
  \hspace{-0.35cm}
  \subfloat{\includegraphics[width=0.335\textwidth]{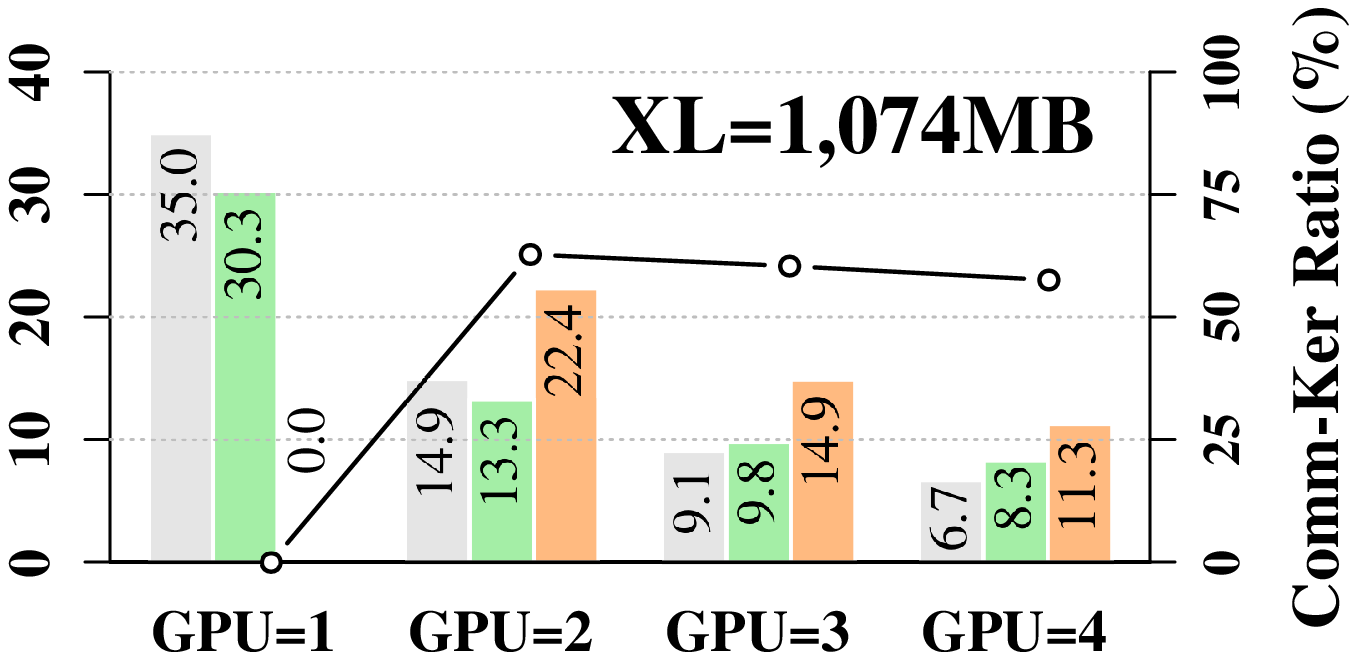}}
  \centering
  \vspace{-10pt}
  \caption{Scaling with the number of GPUs for the stencil kernel in the Himeno benchmark. The grey bar shows the average execution time of a manually-tuned loop-splitting version which includes the communication time and the kernel time. The kernel and communication latencies by our technique are shown by the green and orange bars. The line plots the communication-to-kernel latency ratio. Since the kernel requires only 3D halo accesses, the optimized transfer finishes within 30us, 50us and 110us for the size M, L and XL, respectively, regardless of the number of GPUs used. The displayed data size equals the total grid proportions that all the GPUs update.
  }
  \label{fig:himeno}
  \vspace{-0.2cm}
\end{figure*}

Fig.~\ref{fig:himeno} shows the performance scaling with Himeno using different program sizes.
Since we equally split array ranges for each GPU,
the transfer size per GPU-to-GPU connection becomes smaller 
and the proportion of communication decreases when the number of GPUs is increased.
From two to four-GPU use,
we see different scaling of total GPU-to-GPU transfers:
1.70x speedup with size M, 1.95x with L and 1.99x with XL.
In regard to kernel performance scaling from single to four GPUs,
we achieve 1.53x, 2.19x and 3.64x improvements for size M, L and XL, respectively.
For size M, multi-GPU execution suffers the overheads of both kernel and communication.
Better scaling can be obtained with longer kernel execution and larger transfers
as in the case of BT, which is successfully parallelized with communications of a six-dimensional array decomposed per GPU in 75 segments of 8MB size, having original kernel execution longer than 10ms.

Our technique further reduces the GPU-to-GPU communication latency as more GPUs are used.
As future architectures move to having many accelerators with all-to-all interconnects, applications could benefit further from
predicate-based filtering.

\begin{figure*}[t]
  \vspace*{-0.3cm}
  \centering
  \subfloat{\includegraphics[width=0.335\textwidth]{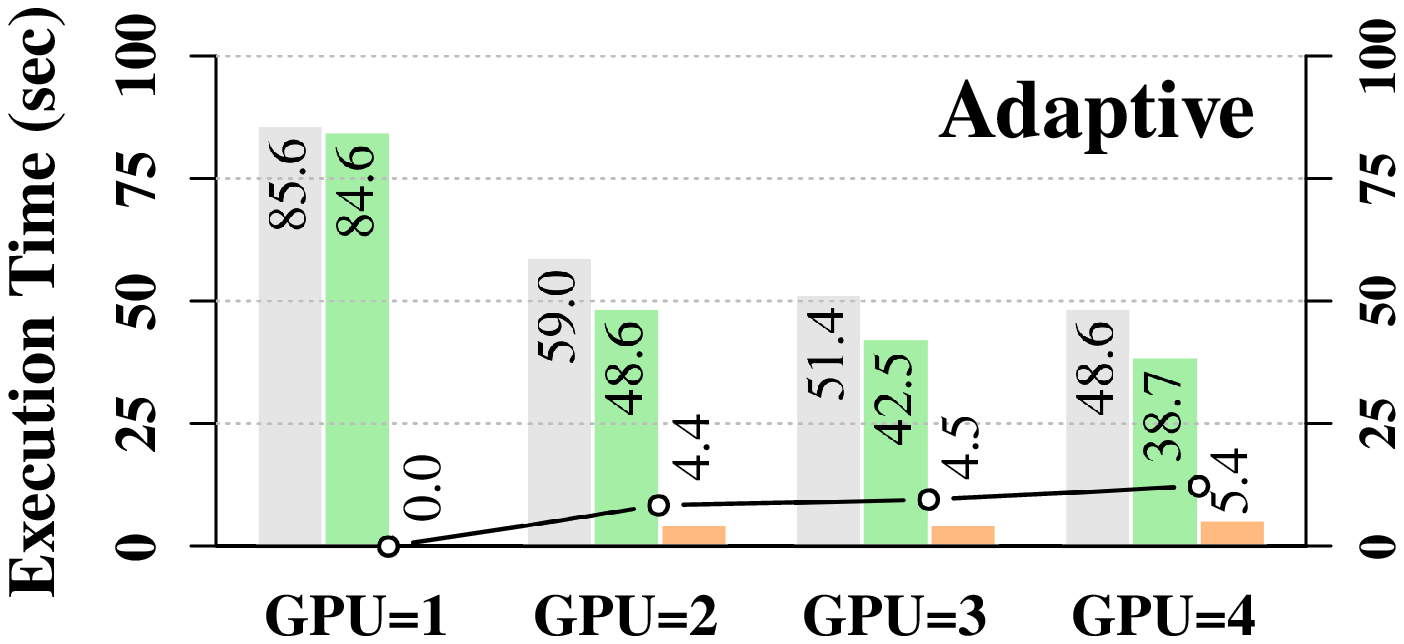}}
  \hspace{-0.36cm}
  \subfloat{\includegraphics[width=0.335\textwidth]{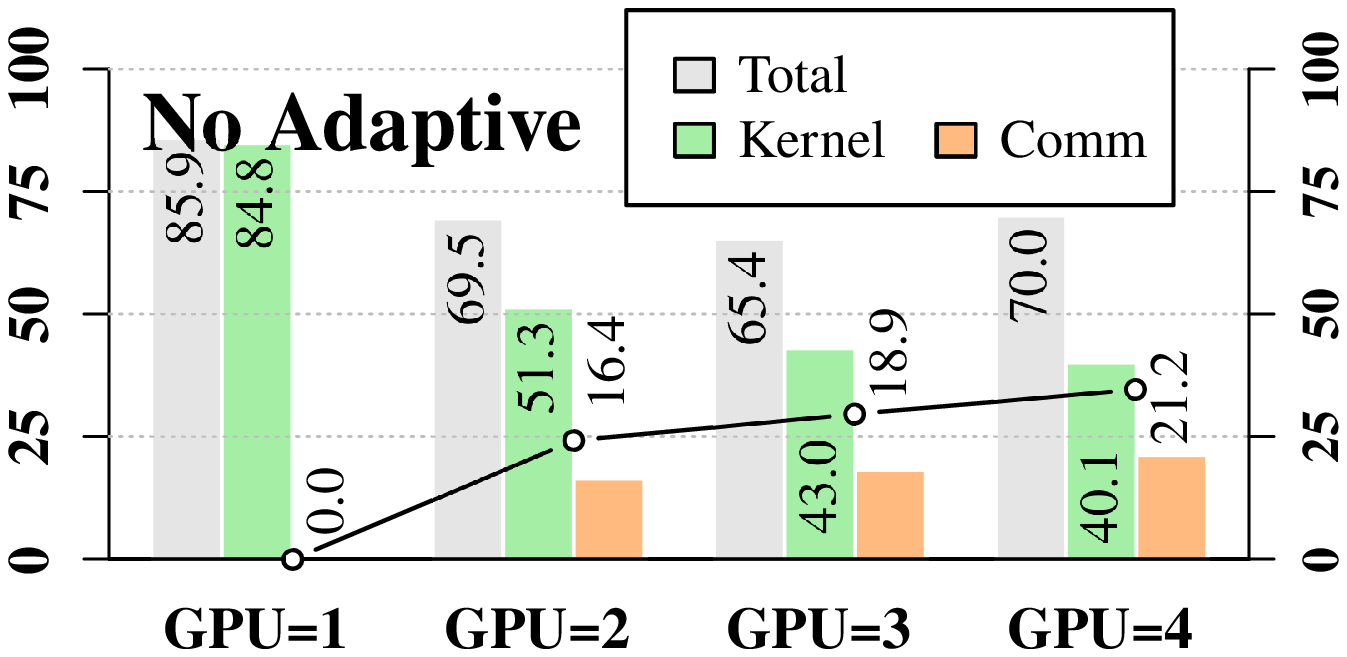}}
  \hspace{-0.35cm}
  \subfloat{\includegraphics[width=0.335\textwidth]{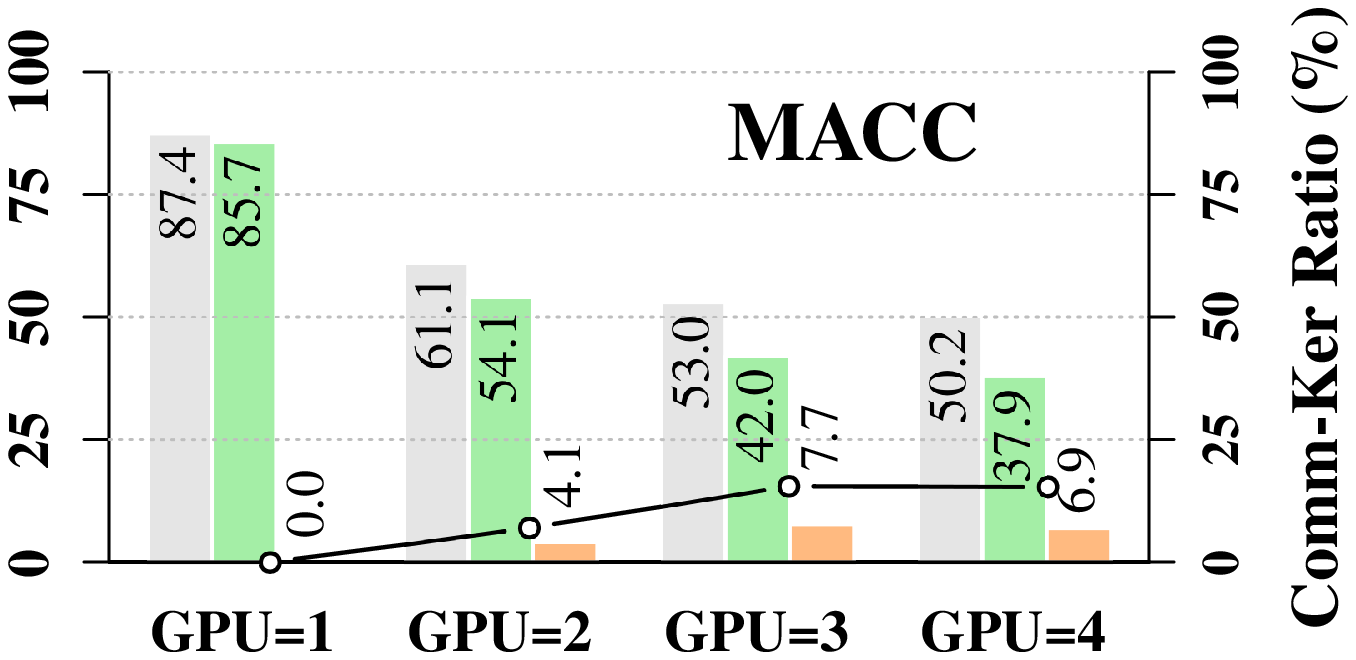}}
  \centering
  \vspace{-10pt}
  \caption{Comparison between predicate-based filtering with/without the adaptive algorithm and MACC using NPB-CG. The total execution, kernel execution and GPU-to-GPU transfer times are shown by the grey, green and orange bars, respectively. The latency ratio of communication to kernel is depicted by the line.}
  \label{fig:cg}
\end{figure*}

\subsubsection{Comparison}

Related work MACC~\cite{10.1007/978-3-319-69953-0_7} successfully parallelizes only two of those applications over multi-GPUs based on code-level access-range analysis: CG and Himeno, which performance bottlenecks have non-overlapping linear writes for each loop iteration. Other multi-GPU work based on memory coherence mechanisms~\cite{komoda2013integrating, distcl, 10.1145/2304576.2304623} is also unable to support the remaining benchmarks without user effort.

We compare our technique to loop splitting.
Fig.~\ref{fig:himeno} includes the scaling of the manual code which uses the same algorithm as MACC.
We notice that the loop-splitting code has better scaling for kernel execution:
from single to four GPUs,
it achieves 2.08x, 3.57x and 5.26x improvements for size M, L and XL, respectively.
Moreover, the optimized communication for the stencil application signif- icantly reduces the latencies.
Those domain-specific approaches can be automated as long as compiler analysis allows; thus, we consider integrating them into our work for future refinement.

Fig.~\ref{fig:cg} shows the comparison between our technique and MACC using CG.
MACC automates loop splitting of all the kernels in CG
but employs no adaptive algorithm.
From two to four-GPU utilization,
our technique achieves better efficiency up to 3.50\%
by disabling multi-GPU execution for lower-latency kernels.
Besides that,
the adaptive execution has smaller kernel latencies than non-adaptive execution
due to the same reason, with 44.09\% better total efficiency.

\section{Related Work}
\label{sec:related}

Several studies conduct optimization upon
the source code of directive-based programming models.
In ~\cite{7573861},
Tian et al. perform scalar replacement on OpenACC code,
that substitutes redundant array accesses with scalar references
until the compiler reports
that all available registers are utilized or all reused
references are replaced.
Barua et al.~\cite{10.1145/3243176.3243196}
optimally
unroll OpenACC loops while estimating memory throughputs based on ILP. 
OptACC~\cite{10.1145/2792745.2792783} finds a better OpenACC parallelism with either grid or direct search.
Hoshino et al.~\cite{hoshino} propose OpenACC directives for array trans- formation.
JACC eases the implementation of those extensional work in a portable fashion to the user's environment while utilizing dynamic information.

There is some work addressing automated multi-GPU utiliza- tion with OpenACC.
MACC~\cite{10.1007/978-3-319-69953-0_7}
provides dynamic access\-/range analysis to
distribute execution with GPU-to-GPU communi- cation.
Although MACC achieves better performance than a UM system,
its analysis is only applicable to affine loops.
Komada et al.'s compiler~\cite{komoda2013integrating} keeps data coherence
by tracking array writes in a similar way to UM but incurring additional array writes for it and performing data transfers after each kernel execution.
Both previous work divides loop execution equally for each GPU
and principally does not allow any intersection of array updates
among devices, thus, cannot support many applications that
our work parallelize.

Distributed-memory systems including multi-GPUs are also dis- cussed
regardless of programming models.
Loop models such as polyhedral model
have been widely employed to detect data dependency
among compute nodes~\cite{1639500, 6877466, 10.1145/2544100, matzautomated}.
However, the input is typically restricted to affine or almost-affine loops
and those work fixes workloads on each node before computing depen- dency,
which involves intricate communication patterns or
im- poses loop transformations beforehand.
Some libraries and frame- works are dedicated to multi-GPU execution
through an abstrac- tion which entails GPU-to-GPU communication~\cite{schaetz2012multi, 10.1145/3018743.3018756, 7832793, nvshmem}.
Software-level memory managements that maintain data coheren- cy are also capable for accommodating program distribution with little user intervention but manual efforts are required to ade- quately partition updated arrays while not overlapping them or otherwise introducing overheads~\cite{10.1145/2581122.2544163,10.1145/2798725,10.1145/2304576.2304623,distcl}.
Our predicate\-/based filtering provides a new way to parallelize many kernels based on source-code level transformation and dynamic infor- mation.

Dynamic compilation
brings additional opportunities for perfor- mance improvement to the runtime system.
NVIDIA's jitifiy~\cite{jitify} is a library that simplifies
the use of CUDA Runtime Compilation (NVRTC).
KernelGen~\cite{6969492} is a Fortran/C compiler that automates GPU code generation with polyhedral loop analysis of LLVM IR.
Those works present dynamic features such as
runtime alias analysis and parameter tuning alongside kernel specialization.
On the other hand, JACC wraps OpenACC compilers and holds C/Fortran code for optimization.

A few projects aim to assist code generation with directive\-/based programming.
Juggler~\cite{10.1145/3178487.3178492} compiles OpenMP task code into a unified GPU program in order to alleviate the latency of global synchronizations.
It requires profiling execution to inspect dy- namic information
and restricts data transfers to be outside task regions.
In Juggler, worker thread-blocks retrieve tasks from queues
while managing the execution with a dependence matrix.
Andi\'{o}n et al.~\cite{10.1007/s10766-015-0362-9} perform program analysis to coalesce data accesses while maximizing register and shared-memory use of directives.
DawnCC~\cite{10.1145/3084540} automates annotating OpenMP/ OpenACC directives based on
static analysis of LLVM IR while coalescing data movements.
CLACC~\cite{8639349} is an OpenACC compiler that converts input into OpenMP.
CCAMP~\cite{10.5555/3433701.3433831} is an interoperable framework for OpenACC and OpenMP, equipped with device\-/specific optimization.
Our dynamic approach complements other work to maximize their scope.

\section{Conclusion}\label{sec:conclusion}

Over the last decade, substantial work has been proposed for
the optimization of directive-based programming models
while facing difficulties with its implementation
to arrange compiler ground- work.
In this paper, we presented JACC, an OpenACC framework which facilitates runtime extension.
Organizing data mappings and kernel arguments as runtime information,
JACC creates dy- namic code from original kernels and compiles it
with a specified compiler in order to support on-the-fly code extension auto- matically.
Due to the memory-bound nature of GPUs,
we proposed predicate-based filtering, a novel code-translation technique of multi-GPU utilization,
for distributing highly-tuned applications without additional user effort.
JACC employs an adaptive algo- rithm for switching distribution
based on the overhead of GPU-to-GPU communication.
Having many kernels parallelized on a multi-GPU environment,
we showed the performance im- provements of several tested benchmarks where precise data\-/dependency analysis is always restrained.

For future work,
we plan to combine predicated-based filtering with other JACC extensions such as automated asynchronous execution and kernel specialization.
Also, static compilation ahead of dynamic compilation is considered for reducing runtime over- heads.
Moreover,
additional program analysis for domain-specific optimization
and aggressive auto-tuning for better efficiency can be incorporated.

\section*{Acknowledgement}

The EPEEC project has received funding from the European Union's Horizon 2020 research and innovation programme under grant~agreement~No~801051.~We~gratefully~acknowledge~the~support\linebreak of NVIDIA Solutions Lab who provided us the remote access to NVIDIA DGX-1.
We would like to acknowledge the NVIDIA AI Technology Center (NVAITC) Europe for their valuable help.

\balance
\bibliographystyle{ACM-Reference-Format}
\bibliography{main}

\end{document}